\documentclass[a4paper,11pt]{article}
\pdfoutput=1 % if your are submitting a pdflatex (i.e. if you have
\usepackage{jheppub} % for details on the use of the package, please
\usepackage[T1]{fontenc} % if needed
\usepackage{natbib}
\usepackage{hyperref}
\usepackage{bbold}

\usepackage{graphicx,amsbsy,amsfonts,amssymb,amsmath}
\usepackage{hyperref}
\usepackage{color}
\usepackage{multirow}
\usepackage{tikz}
\usepackage{stackrel}

\usepackage{booktabs}
\usepackage{braket}
\usepackage{slashed}

\usepackage{adjustbox}

\newcommand{\be}{\begin{equation}}
\newcommand{\ee}{\end{equation}}

\newcommand{\bea}{\begin{eqnarray}}
\newcommand{\eea}{\end{eqnarray}}

\newcommand{\eps}{\epsilon}

\def\sect#1{section~{\ref{#1}}}

\def\eqn#1{eq.~(\ref{#1})}
\def\eqns#1#2{eqs.~(\ref{#1}) and (\ref{#2})}

\def\spa#1.#2{\left\langle#1\,#2\right\rangle}
\def\spb#1.#2{\left[#1\,#2\right]}
\def\spash#1.#2{\spa{\smash{#1}}.{\smash{#2}}}
\def\spbsh#1.#2{\spb{\smash{#1}}.{\smash{#2}}}
\def\sand#1.#2.#3{%
\left\langle\smash{#1}{\vphantom1}^{-}\right|{#2}%
\left|\smash{#3}{\vphantom1}^{-}\right\rangle}
\def\sandpp#1.#2.#3{%
\left\langle\smash{#1}{\vphantom1}^{+}\right|{#2}%
\left|\smash{#3}{\vphantom1}^{+}\right\rangle}
\def\sandpm#1.#2.#3{%
\left\langle\smash{#1}{\vphantom1}^{+}\right|{#2}%
\left|\smash{#3}{\vphantom1}^{-}\right\rangle}
\def\sandmp#1.#2.#3{%
\left\langle\smash{#1}{\vphantom1}^{-}\right|{#2}%
\left|\smash{#3}{\vphantom1}^{+}\right\rangle}

\def\NC{{N_c}}
\def\NF{{N_f}}
\def\CA{\mathcal{A}}
\def\CF{\mathcal{F}}

\DeclareMathOperator{\Tr}{Tr}
\DeclareMathOperator{\sgn}{sgn}

\def\tt{\tilde{t}}

\def\BlackHat{{\sc BlackHat}}
\def\Mathematica{{\sc Mathematica}}

\newbox\charbox
\newbox\slabox
\def\s#1{{      % Feynman slash
        \setbox\charbox=\hbox{$#1$}
        \setbox\slabox=\hbox{$/$}
        \dimen\charbox=\ht\slabox
        \advance\dimen\charbox by -\dp\slabox
        \advance\dimen\charbox by -\ht\charbox
        \advance\dimen\charbox by \dp\charbox
        \divide\dimen\charbox by 2
        \raise-\dimen\charbox\hbox to \wd\charbox{\hss/\hss}
        \llap{$#1$} }}

\title{Planar Two-Loop Five-Parton Amplitudes from Numerical Unitarity}

\author{S. Abreu,$^{a}$ F. Febres Cordero,$^{a,b}$ H. Ita,$^{a}$ B. Page$^{a}$ and  V. Sotnikov$^{a}$}

\affiliation{$^{a}$Physikalisches Institut, Albert-Ludwigs-Universit\"at Freiburg,\\
	Hermann-Herder-Str.~3, D--79104 Freiburg, Germany}
\affiliation{$^{b}$Physics Department, Florida State University,\\
	77 Chieftan Way, Tallahassee, FL 32306, U.S.A.}

\abstract{
We compute a complete set of independent leading-color two-loop
five-parton amplitudes in QCD.
These constitute a fundamental
ingredient for the next-to-next-to-leading order QCD corrections
to three-jet production at hadron colliders.
We show how to consistently consider helicity amplitudes with 
external fermions in dimensional regularization, allowing the
application of a numerical variant of the unitarity method.
Amplitudes are computed by exploiting a decomposition of the 
integrand into master and surface terms that is independent of
the parton type. Master integral coefficients are 
numerically computed in either finite-field or floating-point
arithmetic and combined with known analytic master integrals.
We recompute leading-color two-loop four-parton amplitudes as a
check of our implementation.
Results are presented for all independent four- and five-parton
processes including contributions with massless closed fermion
loops.
}

\begin{document} 
\noindent FR-PHENO-2018-011\hfill
\maketitle
\flushbottom

\newpage

%%%%%%%%%%%%%%%%%%%%%%%%%%%%%%%%%%%%%%%%%%%%%%%%%%

% !TEX root = ../main.tex

\section{Introduction}
\label{Sec:Introduction}

The progress in our understanding of the analytic properties of
loop amplitudes has recently led to the computation of the first
two-loop five-point amplitudes in QCD
\cite{Badger:2013gxa, Badger:2015lda, Gehrmann:2015bfy, 
Dunbar:2016aux, Badger:2017jhb, Abreu:2017hqn}. These
computations focused on the leading-color contributions to the
five-gluon process. 
In this paper we take a further step and compute the
scattering amplitudes of all five-parton processes in the
leading-color limit, including
corrections with massless closed fermion loops.
Two-loop five-parton amplitudes without closed quark loops were
recently presented in ref.~\cite{Badger:2018gip}, and
related work on the complete reduction of two-loop five-parton 
amplitudes appeared in 
refs.~\cite{Boels:2018nrr,Chawdhry:2018awn}.
Our results are an important step towards the automation of
the calculation of two-loop partonic amplitudes, which
are in turn an important ingredient towards obtaining
theory predictions to three-jet production at hadron colliders
at next-to-next-to-leading order (NNLO) in QCD.

The main result of this paper is a numerical method for the
computation of two-loop multi-parton amplitudes, including
massless quark states in addition to gluons. 
We apply a numerical variant of the unitarity method
\cite{Bern:1994zx,Bern:1994cg,Bern:1997sc,Britto:2004nc},
which was extensively used for one-loop computations
\cite{Ossola:2006us,Ellis:2007br,Giele:2008ve,Berger:2008sj} and
recently generalized to two loops 
\cite{Ita:2015tya,Abreu:2017idw,Abreu:2017xsl}.
In this paper we extend its implementation to two-loop processes
involving fermions. 
Four-parton two-loop corrections interfered with their
tree-level amplitudes
\cite{Anastasiou:2000ue,Anastasiou:2000kg,Anastasiou:2001sv,
Glover:2001af} as well as the associated helicity amplitudes
\cite{Bern:2002tk,Glover:2003cm,Bern:2003ck,Glover:2004si,
DeFreitas:2004kmi}
were computed with analytic methods some time ago, and we
recompute these results as a check of our implementation.
The two-loop numerical unitarity method we employ avoids the 
challenging algebra of analytic multi-scale computations and is
at the same time sufficient for the numerical phase-space
integration required in cross-section computations.
To showcase its potential we provide numerical benchmark values
for five-parton amplitudes. To this end we compute integral
coefficients with exact or floating point arithmetic,
and combine them with the numerical evaluation of the two-loop
master integrals \cite{Papadopoulos:2015jft,Gehrmann:2018yef}.
We leave an analysis of the integration over the
physical phase space to future work.

A number of developments is necessary for handling fermions. 
Fermion amplitudes have been computed within the numerical unitarity
method at one loop 
\cite{Ellis:2008ir,Berger:2008sz,Badger:2017gta,Anger:2017glm,
Anger:2018ove}, and in this paper we propose a generalization
for two-loop amplitudes.
We first discuss the treatment of fermion states in
dimensional regularization. At higher loop orders,
the subtleties in this procedure become increasingly relevant 
(see~\cite{Gnendiger:2017pys} for a recent review).
In particular, we discuss in detail the definition of 
dimensionally-regulated {\em helicity} amplitudes with pairs of
external quarks. We present a prescription for how to define a
helicity amplitude at two loops which can be used to compute
interference terms in the 't Hooft-Veltman scheme.
This prescription can be implemented numerically, extending
well known analytic methods \cite{Bern:2002tk,Glover:2003cm,
Bern:2003ck,Glover:2004si,DeFreitas:2004kmi} which are not
directly applicable in a numerical calculation since they rely 
on abstract algebraic manipulations of the $\gamma$-matrix
algebra.

A second technical advance concerns the implementation of the 
numerical unitarity method in exact 
arithmetic~\cite{Abreu:2017hqn} for amplitudes with fermions,
based on the use of finite-fields techniques for amplitude
computations~\cite{Peraro:2016wsq}.
The main obstacle to overcome relates to the fact that generic
polynomial equations do not have solutions in a generic 
algebraic field.
This is in tension with the fact that in a unitarity-based approach
one needs to generate loop momenta satisfying the set of quadratic
on-shell conditions.
We describe a way to handle this difficulty while maintaining the
power of exact finite field computations when considering both gluons
and fermions.
We stress that the ability to perform exact calculations on
rational phase-space points is an additional feature of our
computational framework, and the numerical unitarity method has
also been implemented in more standard floating-point arithmetic.

The article is organized as follows. In 
section~\ref{sec:Methods} we define leading-color 
dimensionally-regulated helicity amplitudes with external
fermions. 
In section~\ref{sec:Implementation} we present explicit details
of our implementation.
In section~\ref{sec:Results}, we first present results for the
leading-color four-parton helicity amplitudes,
and then present our numerical results for a complete set of
independent five-parton amplitudes. We also describe the checks 
we performed.
We give our conclusions and outlook in 
section~\ref{sec:Conclusion}. Finally, we present
some useful $\gamma$-matrix
identities in appendix~\ref{sec:identities}, as well as 
some details on the infrared structure of the amplitudes in
appendix~\ref{sec:IR}.

% !TEX root = ../main.tex

\section{Dimensionally Regulated Helicity Amplitudes}
\label{sec:Methods}

Dimensionally regulated scattering amplitudes are functions of
the continuous dimension parameter $D=4-2\eps$, which
regulates both ultraviolet and infrared singularities of loop
integrals. Within this framework, the state spaces of the
external particles are also naturally understood to be formally
infinite dimensional.
In practice, however, we are interested in obtaining predictions
for physical external states that are strictly four dimensional.
For instance, in this paper we will be computing
helicity amplitudes. 
In order to compute a dimensionally-regulated amplitude with a 
given set of four-dimensional external states, we must find how
to represent them in the $D$-dimensional space. That is, we must
find a consistent embedding of the physical four-dimensional
state in $D$ dimensions. This is trivially achieved for gluon 
helicity states: these
are vector particles and any four-dimensional polarization state
can be embedded in a generic $D$-dimensional space by filling 
the remaining components of the vector with zeros. 
For fermion states, however, the embedding is less trivial as 
the nature of the $D$-dimensional Clifford algebra means that
there is no single associated state in $D$-dimensions.  
As such,
one might wonder if it is possible at all to unambiguously 
define four-dimensional helicity amplitudes with external
fermions. In this section we describe how we
address this problem, inspired by the approach of
refs.~\cite{Glover:2003cm,Glover:2004si}, and precisely define
the objects that we will be computing in subsequent sections.

\subsection{Embedding of Fermionic States in Dimensional
Regularization}
\label{sec:Clifford}

There are several consistent regularization schemes that can be
chosen, see
e.g.~ref.~\cite{Gnendiger:2017pys} for a recent review, and our
discussion applies to both the conventional
dimensional regularization (CDR) and the 't Hooft-Veltman (HV)
schemes. The two schemes differ in the way
vector particles (gluons in our case) are treated,
and we follow 
the description given in the reference
above.\footnote{Except for the meaning of $D_s$, which we use 
to denote the dimension of the CDR and HV singular vector 
fields, to differentiate it from the dimensional regulator $D$.}
In CDR, all vector fields are vectors in a space of dimension
$D_s=4-2\epsilon$. In
HV, one distinguishes between \emph{regular} and 
\emph{singular} vector 
fields. The former do not lead to any singularities and 
are considered to be
strictly four-dimensional objects. The
latter are a source of singularities and are 
$D_s$-dimensional vectors. For our purposes, this means
that gluons whose momentum we integrate over are 
$D_s$ dimensional, and external gluons whose momentum
we do not integrate over are
four dimensional. The two schemes are consistent,
in that their contributions to NNLO computations can be related by known
transition rules~\cite{Broggio:2015dga}, but as we
shall see below the HV scheme introduces some simplification in
the calculation.

We consider fermions in $D_s$ dimensions, as is for instance
necessary when a CDR gluon or a singular HV gluon is emitted 
from a quark line. If the fermion line closes upon itself, as in
e.g.~the $N_f$ corrections to gluon amplitudes
(i.e., corrections with a closed massless-quark loop), 
we only need the defining
property of the Clifford algebra
\begin{equation} \label{eqn:defClifford}
\{ \gamma_{[D_s]}^\mu, \gamma_{[D_s]}^\nu \} = 
2 g_{[D_s]}^{\mu\nu}\mathbb{1}_{[D_s]}^{\vphantom{\mu\nu}} \,,
\end{equation}
where we explicitly write the dimension $D_s$
as a subscript of the $\gamma$-matrices and the metric, and
use a metric with mostly-minus Minkowski signature, 
$g_{[D_s]}={\rm diag}\{1, -1 , ... , -1 \}$.
Here $\mathbb{1}_{[D_s]}$ is the identity operator in the representation space of 
the Clifford algebra.
In the presence of external fermions, however, 
we must also describe the corresponding states and
an explicit
representation of the $D_s$-dimensional Clifford algebra is
required. Since we are ultimately interested in specifying
four-dimensional
external states, it is furthermore convenient to construct 
the representation in a factorized way starting from four dimensions
(see e.g.~refs.~\cite{Kreuzer:susylectures,Collins:1984xc}).
We thus consider a Clifford algebra
in $D_s$ dimensions as the tensor product of a
four-dimensional and a $(D_s-4)$-dimensional one:
\begin{equation}\label{eqn:cliffordrecursion}
(\gamma_{[D_s]}^\mu)_{a\kappa}^{\,b\lambda}  = \left\{ 
\begin{array}{ll} 
		\left(\gamma_{[4]}^\mu\right)_a^{\;b} \,
		\delta_\kappa^\lambda\,, &\quad  0\le\mu \le 3 \,,\\&\\
		\left(\tilde\gamma_{[4]}\right)_a^{\;b} 
		\left(\gamma_{[D_s-4]}^{(\mu-4)}\right)_\kappa^{\;\lambda}\,, 
        &\quad \mu > 3 \,,
\end{array}
		\right.
\end{equation}
where $\tilde\gamma_{[4]}\equiv i(\gamma_{[4]}^0
\gamma_{[4]}^1 \gamma_{[4]}^2 \gamma_{[4]}^3)$, such that 
$(\tilde\gamma_{[4]})^2 =\mathbb{1}_{[4]}$ is the identity
operator
in the four-dimensional algebra.
The indices $a,b$ denote the spinor indices in the
four-dimensional algebra and $\kappa,\lambda$ the ones of the 
$(D_s-4)$-dimensional one.
The $\gamma^\mu_{[D_s-4]}$ form themselves a 
$(D_s-4)$-dimensional Clifford algebra with signature
$g_{[D_s-4]}={\rm diag}\{ -1 , ... , -1 \}$.
In amplitude calculations we naturally encounter products of
$\gamma$ matrices, and in this paper we will mainly focus
on chains of $\gamma_{[D_s-4]}$ matrices. 
We thus define a convenient basis for these chains, 
constructed by anti-symmetrizing over their Lorentz
indices and given by (see e.g.~\cite{Kreuzer:susylectures})
\begin{align}\label{eq:basisGammaChainDs}
\gamma_{[D_s-4]}^{\mu_1 \ldots \mu_n} = \frac{1}{n!} \sum_{ \sigma\in S_n} \sgn(\sigma) \gamma_{[D_s-4]}^{\mu_{\sigma(1)}} \ldots \gamma_{[D_s-4]}^{\mu_{\sigma(n)}}\,,
\end{align}
where $S_n$ denotes the set of all permutations of $n$
integers and $\sgn(\sigma)$ the signature of the permutation
$\sigma\in S_n$.

The spinor states associated with the $D_s$-dimensional 
Clifford algebra live in a $D_t$-dimensional space.\footnote{We 
remind the reader that although $D_t=2^{D_s/2}$ for any 
finite-dimensional representation, $D_t$ is set to $4$ in 
dimensional regularization \cite{Collins:1984xc}.}
For four-dimensional momenta they can be constructed from a 
tensor-product representation as
\begin{equation} 
\label{eq:fermionicStates}
\psi_{s,a \kappa} =  (u_h)_a (\eta^i)_\kappa \,,
\quad\mbox{and}\quad
\bar \psi_s^{a \kappa} = 
(\bar u_h)^{a}  (\bar \eta_i)^\kappa\,,
\end{equation}
where we have introduced an index $s=\{h, i \}$ to denote the
polarization states in terms of spinors of the four- and 
$(D_s-4)$-dimensional subspaces. 
Without loss of generality we can
require that $(\eta^i)_\kappa$ and $(\bar \eta_i)^\kappa$ 
be dual to each other,
\begin{equation}
\label{eqn:qspinors}
(\bar \eta_i)^\kappa (\eta^j)_\kappa = \delta_i^j\,,
\end{equation}
and choose a canonical basis for the 
spinors in the $(D_s-4)$-dimensional space,
i.e.~set $(\eta^i)_\kappa=\delta^i_\kappa$.
In an on-shell computation in $D_s$ dimensions we use the 
spinor states defined in eq.~\eqref{eq:fermionicStates}
as external fermion wave functions.
Given the choice of a canonical basis for the 
$(D_s-4)$-dimensional states, we can identify the
$(D_s-4)$ polarization label $i$ with the spinor index 
$\kappa$ in eq.~\eqref{eq:fermionicStates}.
Thus, in the following we only insert four-dimensional spinors
and keep track of the $(D_s-4)$-dimensional embedding with 
the open $(D_s-4)$ spinor index.
We note that a bilinear of external $(D_s-4)$-dimensional
spinors $\eta \bar \eta'$ can  be expressed in terms of the 
basis of $\gamma$-matrix chains introduced in 
eq.~\eqref{eq:basisGammaChainDs},
\begin{equation}\label{eq:etaetaP}
\eta \bar \eta' = \mathbb{1}_{[D_s-4]} f +  
\gamma_{[D_s-4]}^{\nu_1} f_{\nu_1} + 
\gamma_{[D_s-4]}^{\nu_1\nu_2} f_{\nu_1\nu_2} + \ldots \,,
\end{equation}
where the $\{ f , f_{\nu_1}, f_{\nu_1\nu_2},\cdots \}$ are the
constant coefficients in the decomposition.
In loop calculations, this naively introduces reference vectors and tensors that can yield linear
dependence on the components of the loop-momenta beyond four
dimensions, see e.g.~\cite{Badger:2017gta,Anger:2018ove},
in contrast with
what happens, for instance, in the case of amplitudes with
only gluons.
We only mention this here as an observation since, as we 
shall see in the remaining of this paper, this will not be an
issue with our definition of helicity amplitudes.

The tensor product representation of the Clifford algebra is
particularly useful to separate four- and 
$(D_s-4)$-dimensional spinor indices in $\gamma$-matrix chains. 
Indeed, a product of
$\gamma$ matrices where some
Lorentz indices are within four dimensions (denoted
$\mu_i$),
and some are beyond four dimensions (denoted $\hat\mu_i$) is
split into two blocks, a four-dimensional and a 
$(D_s-4)$-dimensional one. For instance, we have
\begin{align}\begin{split}
\left(\gamma_{[D_s]}^{\mu_1}
\gamma_{[D_s]}^{\hat \mu_2}
\gamma_{[D_s]}^{\mu_3}
\gamma_{[D_s]}^{\hat\mu_4}\right)_{a\kappa}^{\,b\lambda} &=
-\left(
\gamma_{[D_s]}^{\mu_1}
\gamma_{[D_s]}^{\mu_3}
\gamma_{[D_s]}^{\hat \mu_2}
\gamma_{[D_s]}^{\hat\mu_4} \right)_{a\kappa}^{\,b\lambda} \\
&=-\left( 
\gamma_{[4]}^{\mu_1}
\gamma_{[4]}^{\mu_3}
\right)_a^{\,b}   \,\left(
\gamma_{[D_s-4]}^{(\hat \mu_2-4)}
\gamma_{[D_s-4]}^{(\hat \mu_4-4)}
\right)_\kappa^{\,\lambda}  \,.
\end{split}\end{align}
Consider now contracting the above
product of $\gamma$ matrices with a four-dimensional fermion state,
such as the $u$ and $\bar u$ spinors:
\begin{equation}
	\label{eq:tensorChain}
	\bar u^a
	\left(\gamma_{[D_s]}^{\mu_1}
	\gamma_{[D_s]}^{\hat \mu_2}
	\gamma_{[D_s]}^{\mu_3}
	\gamma_{[D_s]}^{\hat\mu_4}\right)_{a\kappa}^{\,b\lambda}
	u_b=
	-\left(\bar u
	\gamma_{[4]}^{\mu_1}
	\gamma_{[4]}^{\mu_3}
	u\right)
	\left(
	\gamma_{[D_s-4]}^{(\hat \mu_2-4)}
	\gamma_{[D_s-4]}^{(\hat \mu_4-4)}
	\right)_\kappa^{\,\lambda}\,.
\end{equation}
The result is a tensor with open indices in the
$(D_s-4)$-dimensional space.
We recall that these are in one-to-one
correspondence with a $(D_s-4)$-dimensional state, and the above expression (\ref{eq:tensorChain}) is
thus equivalent to a contraction with on-shell helicity states in $D_s$ dimensions.
Our ultimate goal is the calculation of amplitudes relevant for 
cross-section computations, and we must then understand which
tensor structures beyond four dimensions are necessary. 
This will be done in the next subsections.

We finish this section with two comments. First, we note that in the
HV scheme this tensor decomposition results in simpler expressions
than in CDR. Consider for instance the tree-level $q\bar q\to Q\bar
Q$ amplitude
\begin{equation}
	\left(M^{(0)}\right)_{\kappa_1 \kappa_2}^{\lambda_1 \lambda_2}
	\;\sim\;\; 
        \bar u^{a_1}
	\left(
	\gamma_{[D_s]}^\mu
	\right)_{a_1 \kappa_1}^{b_1 \lambda_1}
	u_{b_1}
        \,\,
	\bar u^{a_2}
	\left(
	\gamma_{[D_s]\mu}
	\right)_{a_2 \kappa_2}^{b_2 \lambda_2}
	u_{b_2}
	\,.
\end{equation}
In the HV scheme, the gluon between the two quark lines is
four dimensional, i.e., $\mu\leq3$, while
in the CDR scheme, the gluon is $D_s$ dimensional.
From eq.~\eqref{eqn:cliffordrecursion} we thus get
\begin{equation}\label{eq:HVvsCDR}
	\left(M^{(0)}\right)_{\kappa_1,\kappa_2}^{\lambda_1,\lambda_2}
	=\begin{cases}
	M^{(0)}_{0}\, \delta_{\kappa_1}^{\lambda_1}
	\delta_{\kappa_2}^{\lambda_2}& \textrm{in HV,}\\
	M^{(0)}_{0}\, \delta_{\kappa_1}^{\lambda_1}
	\delta_{\kappa_2}^{\lambda_2}+
	M^{(0)}_{1}\,\Bigl(\gamma_{[D_s-4]}^{\mu}\Bigl)_{\kappa_1}^
	{\lambda_1}
	\Bigl(\gamma_{[D_s-4]\mu}\Bigl)_{\kappa_2}^{\lambda_2}
	&\textrm{in CDR,}
	\end{cases}
\end{equation}
where the $M_i^{(0)}$ are coefficients that are determined from
products of four-dimensional $\gamma$-matrices contracted with
four-dimensional spinors.
In the HV scheme the amplitude is
determined by a single coefficient, 
while in CDR two are needed. In the
remainder of this paper we thus choose to work in the HV scheme.
Nevertheless, our discussion
generalizes to the CDR scheme in a straightforward way.

The second comment we wish to make is that, 
although in this section we consider $D_s=4-2\epsilon$, which
means the Clifford algebra defined in 
eq.~\eqref{eqn:cliffordrecursion} is infinite dimensional, in
numerical calculations one might need to construct an explicit
representation of the Clifford algebra and thus
take $D_s$ to be an even integer (larger than 4).
The construction 
of eq.~\eqref{eqn:cliffordrecursion} still holds and, in fact, 
it can be iterated: any even $D_s$ can be reached
by constructing a tensor product of the
$(D_s-2)$ algebra with a 2-dimensional algebra, even if the 
$(D_s-2)$ algebra 
was already constructed as a tensor product of two
algebras.

\subsection{Tensor Decomposition of Helicity Amplitudes}
\label{sec:HelAmplHV}

We consider a helicity amplitude $M$, expanded in 
perturbation theory, with the $k$-th order term written as
$M^{(k)}$.
We saw previously that these are tensors in the
$(D_s-4)$-dimensional spinor space, see
eq.~\eqref{eq:HVvsCDR} for an explicit example.
Here we introduce a basis for the associated tensor space in the
spinor indices beyond four dimensions, whose elements are 
denoted as $v_n$. In general,
the basis depends on the physical process described 
by $M$ and on the order $k$ in the perturbative
expansion. We will suppress this dependence for simplicity of 
the notation and write
\begin{equation}
  M^{(k)}= \sum_n  v_n M^{(k)}_n,
\label{eq:tensorDecomposition}
\end{equation}
where the $M^{(k)}_n$ are computed from
$\gamma_{[4]}^{\mu}$ matrices and external states in four dimensions, and the tensor structure
of the amplitude in the spinor indices beyond four dimensions is
fully contained in the $v_n$.
In the following, we explicitly construct the basis $\{v_n\}$
for two families of amplitudes: those with a pair $q\bar q$ of
external
quarks and any number of external gluons, and
those with two pairs $q\bar q$ and $Q\bar Q$
of external quarks (of either different or identical
flavor) and any number of external gluons.

The different tensors $v_n$ are constructed by contracting the
Lorentz indices of chains of $\gamma_{[D_s]}$ matrices with
other Lorentz vectors in the amplitude after all loop
integrations have been performed. The remaining objects that
carry Lorentz indices are four-dimensional external momenta,
four-dimensional polarization vectors and chains of $\gamma_{[D_s]}$
matrices. Any Lorentz index in a $\gamma_{[D_s]}$-matrix 
chain that is contracted with a four-dimensional object becomes
four-dimensional, contributing only a trivial tensor structure 
in the $(D_s-4)$-dimensional space. For instance, if $\varepsilon_\mu$ represents a
four-dimensional polarization vector of an external gluon,
\begin{equation}
	\label{eq:trivialTens1}
	\varepsilon_\mu
	\left(
	\gamma_{[D_s]}^\mu
	\right)_{a\kappa}^{b\lambda}
	=\varepsilon_\mu\left(\gamma_{[4]}^\mu
	\right)_{a}^{b}\delta_{\kappa}^{\lambda}\,.
\end{equation}
Similarly when two Lorentz indices are contracted inside the 
same chain of $\gamma_{[D_s]}$ matrices, the tensor structure
beyond four dimensions is trivial, as follows from:
\begin{equation}
	\label{eq:trivialTens2}
	\left(\gamma_{[D_s]}^\mu\right)_{a \kappa}^{b_1 \lambda_1}
        \left(\gamma_{[D_s]\mu}^{\phantom{\mu}}\right)_{b_1 \lambda_1}^{b \lambda}
	=D_s\delta_{a}^{b}\delta_{\kappa}^{\lambda}\,.
\end{equation}
Non-trivial tensors $v_n$ are obtained by contracting
Lorentz indices of two chains of $\gamma_{[D_s-4]}$ matrices.
The basis introduced in eq.~\eqref{eq:basisGammaChainDs} 
for these chains is particularly useful for
computing these contractions.

Let us consider an amplitude with a pair $q\bar q$ of external
quarks and any number of external gluons. There is a single
chain of $\gamma_{[D_s-4]}$ matrices and,
as there are no other objects with $(D_s-4)$ indices, it
follows from the discussion above that
for this case there is a single term in the sum of
eq.~\eqref{eq:tensorDecomposition}:
\begin{equation}\label{eq:decompqqbar}
	M^{(k)}
	(q,\bar q,g,\ldots,g)
	=w_0\,M^{(k)}_0\,,\qquad
	\textrm{with}\quad
	(w_0)_{\kappa}^{\lambda}=\delta_{\kappa}^{\lambda}\,.
\end{equation}
We define the dual tensor
$w^0$ such that $w_0\cdot w^0=1$, with more details given in
appendix~\ref{sec:identities}.

Let us now consider an amplitude with two quark pairs of 
different flavors, $q\bar q$ and $Q\bar Q$, and any number of
gluons. We can now contract Lorentz indices between two 
different chains of $\gamma$ matrices, and the basis 
$\{v_n\}$ is then larger in this case. Using the basis for the
$\gamma$-matrix chains introduced in 
eq.~\eqref{eq:basisGammaChainDs}, we construct the associated
basis $\{v_n\}$:
\begin{align}\begin{split} \label{eqn:4qtensors}
  (v_0)_{\kappa_1\kappa_2}^{\lambda_1\lambda_2}   = &
  \delta_{\kappa_1}^{\lambda_1} \delta_{\kappa_2}^{\lambda_2}\,, \\
  (v_1)_{\kappa_1\kappa_2}^{\lambda_1\lambda_2}=
  &(\gamma_{[D_s-4]}^{\mu_1} )_{\kappa_1}^{\lambda_1} 
  (\gamma_{[D_s-4]\mu_1}^{\phantom{\mu}})_{\kappa_2}^{\lambda_2}\,, \\
  & \vdots\\
  (v_m)_{\kappa_1\kappa_2}^{\lambda_1\lambda_2}=
  &(\gamma_{[D_s-4]}^{\mu_1 \ldots  \mu_m})_{\kappa_1}^{\lambda_1}
  (\gamma_{[D_s-4]\mu_1 \ldots \mu_m}^{\phantom{\mu}})_{\kappa_2}^{\lambda_2}\,,\\
  & \vdots\,
\end{split}\end{align}
where we have made explicit the indices in the $
(D_s-4)$-dimensional space.
The basis $\{v_n\}$ is infinite dimensional for
$D_s=4-2\epsilon$ (because there are infinitely many independent
terms of the form of eq.~\eqref{eq:basisGammaChainDs}), but at
each order in the perturbative
expansion only a finite number of basis elements contribute, 
as follows from inspecting the corresponding Feynman diagrams.
We thus have
\begin{equation} \label{eqn:4qampltensor}
	M^{(k)}
	(q,\bar q,Q,\bar Q,g,\ldots,g)
	=\sum_{n=0}^{n_k}
    v_n
	M^{(k)}_n\,.
\end{equation}
In the HV scheme, the decomposition is independent of the number
of external gluons. In particular,
the value of $n_k$ can be determined from the amplitude with no 
external gluons, by examining the Feynman diagrams
with the most singular gluons. 
These are ladder-type four-point diagrams with the 
gluons in the rungs. We find for instance that $n_0=0$,
$n_1=3$ and $n_2=5$ for tree-level, one- and two-loop
amplitudes, respectively. Our decomposition is similar to the 
one presented in ref.~\cite{Glover:2004si}, but differs in the
choice of basis tensors in eq.~\eqref{eq:tensorDecomposition}.

In practical calculations, one is interested in computing
specific coefficients $M^{(k)}_n$ in the decomposition
of eq.~\eqref{eq:tensorDecomposition}. We construct the
basis $\{v_n\}$ such that this operation is trivial, i.e., it
satisfies
\begin{equation}\label{eqn:vnproducts}
	v_n^\dagger\cdot v_m=c_n\delta^n_m\,,
	\quad
    c_0(D_s) = 1\,\quad\mbox{and}\quad  c_{n>0}(D_s)={\cal O}(\epsilon)\,.
\end{equation}
The calculation of the coefficients $c_n$ requires some 
technical operations on $\gamma$ matrices that we present in
appendix~\ref{sec:identities}.
We then construct the dual basis $\{v^n\}$, with elements
\begin{equation}\label{eq:dualBasis}
	v^n=\frac{1}{c_n} (v_n)^\dagger\,.
\end{equation}
Using the dual basis, we directly get
\begin{equation}\label{eqn:helampl}
  M^{(k)}_n = v^n \cdot M^{(k)}.
\end{equation}

Finally let us consider an amplitude with two identical quark
pairs, which can be constructed by anti-symmetrizing the
distinct-flavor amplitude $M^{(k)}$ over the two
flavors~\cite{DeFreitas:2004kmi,Glover:2004si}.
It is then easy to see that the decomposition of
eq.~\eqref{eq:tensorDecomposition} requires an enlarged basis
compared to the distinct-quark case of
eq.~\eqref{eqn:4qtensors}. We thus define the tensors
$\{\tilde v_n\}$ as
\begin{equation}\label{eq:basisIdentical}
	(\tilde v_n)_{\kappa_1\kappa_2}^{\lambda_1\lambda_2}=
	(v_n)_{\kappa_1\kappa_2}^{\lambda_2\lambda_1}\,,
\end{equation}
and the decomposition of eq.~\eqref{eq:tensorDecomposition}
is over the sets $\{v_n\}$ and $\{\tilde v_n\}$.
The basis tensors satisfy
\begin{equation}\label{eq:dualBasisIdentical}
	v_nv^m=\delta^m_n\,,\qquad
	\tilde v_n\tilde v^m=\delta^m_n\,,\qquad
	v_n\tilde v^m=
	\delta^m_0\,\delta_{n,0}+\mathcal{O}(\epsilon)\,,
\end{equation}
where the set $\{\tilde v^n\}$ is constructed to be dual to
$\{\tilde v_n\}$ in the same way as in 
eq.~\eqref{eq:dualBasis}.

We finish this subsection with a comment on the case where $D_s$
is a finite integer $D_s^0$. 
All the discussion above holds, but one
must be careful with a small detail. 
The basis of the Clifford algebra in 
eq.~\eqref{eq:basisGammaChainDs} now contains only a finite 
number of terms, and the basis of tensors $\{v_n\}$ is
consequently restricted by the dimension $D^0_s$. 
If one wants to compute the coefficient of a given tensor 
$v_i$, one must thus choose $D_s^0$ large enough such
that $v_i \in \{v_n\}$. Nevertheless, one can check that a 
calculation done in $D_s^0$ dimensions agrees 
with the $D_s=D_s^0$ limit of the same calculation done in
generic $D_s$.

\subsection{Two-loop Helicity Amplitudes for NNLO Phenomenology}

We have established that helicity amplitudes in dimensional
regularization are tensors in the $(D_s-4)$-dimensional space and
introduced a basis of that space on which we can decompose the
amplitude.  We should in
principle compute all coefficients in the decomposition. However, it
turns out that in a given phenomenological application not all
coefficients may be relevant. We discuss below the two cases
involving external quarks pertinent to the subject of this 
paper, the amplitudes with only external gluons being trivial in
this regard.

\paragraph{Two-loop $q\bar qg\ldots g$ amplitude:}
For the case of an amplitude with a pair $q\bar q$ of external
quarks and any number of external gluons,
there is a single coefficient to determine, 
see eq.~\eqref{eq:decompqqbar}. At order $k$ in perturbation
theory we call this object $A^{(k)}$. It is
computed using
\begin{equation}\label{eq:qqProj}
    A^{(k)}(q,\bar q,g,\ldots,g)=M^{(k)}_0(q,\bar q,g,\ldots,g) =
	w^0\cdot M^{(k)}(q,\bar q,g,\ldots,g)\,,
\end{equation}
i.e. by
tracing over the $(D_s-4)$-dimensional indices of the
fermion line. In this paper we are mostly interested in $k=2$.

\paragraph{Two-loop $q\bar qQ\bar Qg\ldots g$ amplitude:}
For a
two-loop amplitude with two quark pairs of 
different flavors, $q\bar q$ and $Q\bar Q$, and any number of
gluons there are in principle six coefficients to determine. 
However, in an NNLO computation (that is not loop-induced) 
the two-loop amplitude is 
interfered with the tree amplitude, which has a single tensor
structure in the HV scheme. The contribution we must compute is of the form
\begin{equation}\label{eq:qqQQProj}
  \left(M^{(0)}\right)^\dagger M^{(2)}=
  \left(M^{(0)}_0\right)^\dagger M^{(2)}_0\,,
\end{equation}
where we have used the orthogonality of the tensors
$v_n$ and the fact that $c_0(D_s)=1$, see 
eq.~\eqref{eqn:vnproducts}. 
For NNLO corrections, it is
thus sufficient to compute the coefficients
$M^{(2)}_0$ through
\begin{equation}
  \label{eq:AmplitudeDefinition}
  A^{(2)}(q,\bar q,Q,\bar Q,g,\ldots,g)=
  M^{(2)}_0 = v^0 \cdot 
  M^{(2)}(q,\bar q,Q,\bar Q,g,\ldots,g),
\end{equation}
which amounts to computing the 
$(D_s-4)$-dimensional trace of $M^{(2)}$ on
each fermion line.  
We define the amplitude
$A^{(k)}(q,\bar q,Q,\bar Q,g,\ldots,g)$ for any order $k$
in an analogous way.

This approach is similar to the one of
ref.~\cite{Glover:2004si} and is in
agreement with the prescription of ref.~\cite{Anger:2018ove}.
On a first look, it might however look inconsistent with the 
way $q\bar qQ\bar Q$ helicity amplitudes are defined in
ref.~\cite{DeFreitas:2004kmi}. Written in the formalism
we have introduced in this section, the authors compute
\begin{equation}
	\label{eq:AmpDefAlt}
	\tilde v^0\cdot M^{(2)}
	(q,\bar q,Q,\bar Q)\,,
\end{equation}
and, given the relations of eq.~\eqref{eq:dualBasisIdentical}, this
would not necessarily give the same 
$A^{(2)}(q,\bar q,Q,\bar Q)$ defined in 
eq.~\eqref{eq:AmplitudeDefinition}. For
phenomenological applications, however, 
one can show that only the
so-called \emph{finite remainder} is relevant 
\cite{Weinzierl:2011uz},
and we now show that the choices of 
eqs.~\eqref{eq:AmplitudeDefinition} and \eqref{eq:AmpDefAlt}
give the same result for this quantity.\footnote{This was 
already pointed out by the authors of 
ref.~\cite{DeFreitas:2004kmi}, who discuss the agreement of 
their finite remainder results with those of ref.~\cite{Glover:2004si}.}
We first recall that the infrared poles of a renormalized QCD
amplitude $M_R$ have a universal structure, and we can
write an amplitude in terms of its universal pole structure
and a finite remainder
which we will denote $\CF$
\cite{Catani:1998bh,Sterman:2002qn,Becher:2009cu,Gardi:2009qi}.
More explicitly, 
for a two-loop amplitude we have
\begin{equation}\label{eq:remainderDef}
	M^{(2)}_R=
	{\bf I}^{(2)}M^{(0)}_R
	+{\bf I}^{(1)}M^{(1)}_R
	+\CF^{(2)}\,,
\end{equation}
where ${\bf I}^{(1)}$ and ${\bf I}^{(2)}$ are operators in color
space. We refer the reader to appendix~\ref{sec:IR} for 
explicit expressions for these operators in the leading-color
approximation of the amplitudes considered in this paper. Since
$\CF^{(2)}$ is finite, we have
\begin{equation}
	v^0\cdot \CF^{(2)}
	=\tilde v^0\cdot \CF^{(2)}+\mathcal{O}(\epsilon),
        \label{eq:RemainderEquivalence}
\end{equation}
and the remainder computed from 
eq.~\eqref{eq:AmplitudeDefinition} thus agrees with the 
one computed from eq.~\eqref{eq:AmpDefAlt}. 

Finally, we now show that in the case of two pairs of
identical quarks we can also use the
definition of eq.~\eqref{eq:AmplitudeDefinition} 
for NNLO phenomenology.
The relevant contribution is the interference of the 
tree-level amplitude with the remainder, i.e.
\begin{equation}
  \left(M^{(0)} - \tilde{M}^{(0)}\right) \cdot \left( \CF^{(2)} - \tilde{\CF}^{(2)}\right)
  =
  \left(M^{(0)}_0 - \tilde{M}^{(0)}_0 \right) \left( v_0\cdot \CF^{(2)} - \tilde{v}_0\cdot \tilde{\CF}^{(2)}\right) + O(\eps),
  \label{eq:LikeInterference}
\end{equation}
where we denote with tildes the flavor exchanged objects.  Here, we
have used the orthogonality of the $v_n$ and $\tilde{v}_n$ up to
$O(\eps)$ to simplify the expression. Importantly, the right hand side
of eq.~\eqref{eq:LikeInterference} now only contains terms that can be
computed through the definition of 
eq.~\eqref{eq:AmplitudeDefinition}.

\subsection{Leading-Color Amplitudes}
\label{sec:lcAmps}

In this paper we compute a complete set of 
independent four- and five-parton helicity amplitudes in the
leading-color approximation. More concretely, we keep the 
leading terms in the formal limit of a large number of colors
$\NC$, and scale the number of massless flavors $\NF$ 
whilst keeping the ratio $\NF/\NC$ fixed. 
Each amplitude can be decomposed in terms of color structures
whose coefficients are related by symmetry, and in this section
we define our notation for the color decomposition of the
amplitudes. We denote the fundamental generators of the
$SU(\NC)$ group by $(T^a)^{\;\bar{\jmath}}_{i}$, where the
adjoint
index $a$ runs over $\NC^2-1$ values and the 
(anti-) fundamental indices $i$ and $\bar \imath$
run over $\NC$ values. We use the normalization 
$ \Tr(T^a T^b) = \delta^{ab}$.

In this work, we will compute amplitudes where the external 
partons have well defined (either positive or negative) 
helicities, following the conventions of 
ref.~\cite{Maitre:2007jq}.
We first discuss the four-point amplitudes. We will consider 
amplitudes for the scattering of four gluons, one quark pair and
two gluons, and
two distinct quark pairs. In the leading-color approximation we
write
\begin{align}
  \begin{split}
  \label{eq:bareAmp4p}
  A(1_g, 2_g, 3_g, 4_g) \big\vert_{\textrm{leading color}} = &
  \sum_{\sigma\in S_4/Z_4} \Tr\left(
  T^{a_{\sigma(1)}} T^{a_{\sigma(2)}} 
  T^{a_{\sigma(3)}} T^{a_{\sigma(4)}} \right)\\
  &\times \CA({\sigma(1)}_g, {\sigma(2)}_g, {\sigma(3)}_g, {\sigma(4)}_g)\,, 
  \end{split} \\
  \begin{split}
  \label{eq:ColorDec2Q4p}
  A(1_q, 2_{\bar{q}}, 3_g, 4_g) \big\vert_{\textrm{leading color}} 
    = & \sum_{\sigma\in S_2} 
  \left( T^{a_{\sigma(3)}} T^{a_{\sigma(4)}} \right)^{\;\bar{\imath}_2}_{i_1} \\
  & \times\CA(1_q,2_{\bar{q}},\sigma(3)_g,\sigma(4)_g)\,,
  \end{split}\\
  \begin{split}
  \label{eq:ColorDec4Q4p}
  A(1_q, 2_{\bar{q}}, 3_Q, 4_{\bar{Q}})
  \big\vert_{\textrm{leading color}} 
  = & 
  \,\delta^{\;\bar{\imath}_{2}}_{i_{3}} \delta^{\;\bar{\imath}_
  {4}}_{i_{1}}\;
  \CA(1_{q}, 2_{\bar{q}}, 3_{Q}, 4_{\bar{Q}}) \,,
\end{split}
\end{align}
where $S_n$ denotes all permutations of $n$ indices and $S_n/Z_n$ denotes all
non-cyclic permutations of $n$ indices. 
We write the particle type explicitly as a subscript, and all
remaining properties of each particle (momentum, helicity, etc.) 
are implicit in the associated number. In the case of
amplitudes involving quarks, we recall that the amplitudes $A$
have been defined in eqs.~\eqref{eq:qqProj} and 
\eqref{eq:AmplitudeDefinition}.
For the five-point case, we will consider the 
amplitudes for the scattering of five gluons, 
one quark pair and three gluons, and
two distinct quark pairs and one gluon. 
In the leading-color approximation we write
\begin{align}
  \begin{split}
  \label{eq:bareAmp}
  A(1_g, 2_g, 3_g, 4_g, 5_g) \big\vert_{\textrm{leading color}} = &
  \sum_{\sigma\in S_5/Z_5} \Tr\left(
  T^{a_{\sigma(1)}} T^{a_{\sigma(2)}} 
  T^{a_{\sigma(3)}} T^{a_{\sigma(4)}} T^{a_{\sigma(5)}} \right)\\
  &\times \CA({\sigma(1)}_g, {\sigma(2)}_g, {\sigma(3)}_g, {\sigma(4)}_g, {\sigma(5)}_g)\,, 
  \end{split} \\
  \begin{split}
  \label{eq:ColorDec2Q}
  A(1_q, 2_{\bar{q}}, 3_g, 4_g, 5_g) \big\vert_{\textrm{leading color}} 
    = & \sum_{\sigma\in S_3} 
  \left( T^{a_{\sigma(3)}} T^{a_{\sigma(4)}} T^{a_{\sigma(5)}} \right)^{\;\bar{\imath}_2}_{i_1} \\
  & \times\CA(1_q,2_{\bar{q}},\sigma(3)_g,\sigma(4)_g,\sigma(5)_g)\,,
  \end{split}\\
  \begin{split}
  \label{eq:ColorDec4Q}
  A(1_q, 2_{\bar{q}}, 3_Q, 4_{\bar{Q}}, 5_g)
  \big\vert_{\textrm{leading color}} 
  = & 
  \,(T^{a_5})^{\;\bar{\imath}_{2}}_{i_{3}} \delta^{\;\bar{\imath}_{4}}_{i_{1}}\;
  \CA(1_{q}, 2_{\bar{q}}, 5_g, 3_{Q}, 4_{\bar{Q}}) \,\,+  \\
  & \,(T^{a_5})^{\;\bar{\imath}_{4}}_{i_{1}} \delta^{\;\bar{\imath}_{2}}_{i_{3}}\;
\CA(1_{q},2_{\bar{q}},3_{Q},4_{\bar{Q}},5_g) \,,
\end{split}
\end{align}
with similar notation as in the four-point case.
For both the four- and five-point cases,
the amplitude with two identical quark pairs can be obtained by
anti-symmetrizing over the distinct flavors 
$q$ and $Q$ as discussed in the previous subsection.

\begin{figure}[]
  \begin{center}
    \begin{tikzpicture}[scale=.9]
    % 5 point masters
    \node at (0,0){\includegraphics[scale=0.5]
    {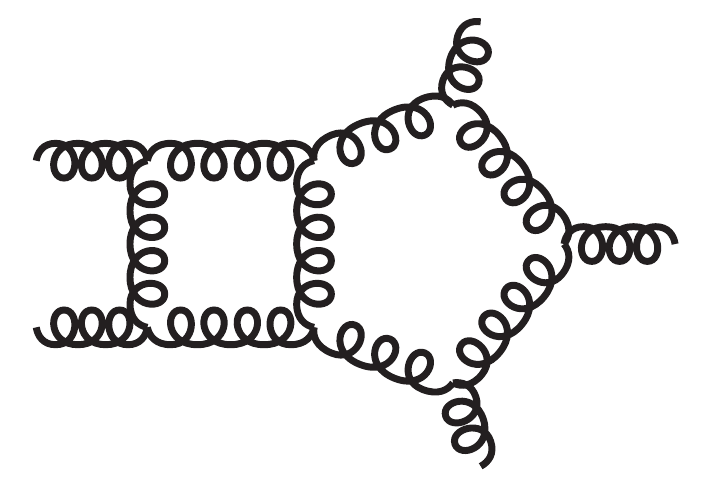}};
    % 5 point masters
    \node at (5,0){\includegraphics[scale=0.5]
    {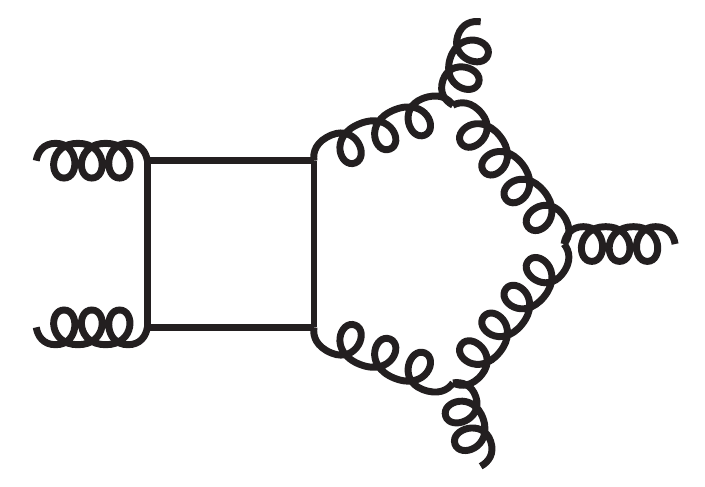}};
    % Level 2 
    \node at (10,0){\includegraphics[scale=0.5]
    {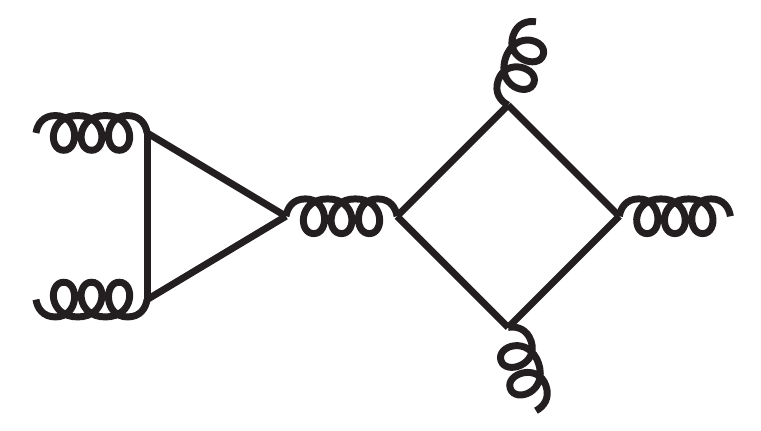}};
\end{tikzpicture}
\end{center} 
\caption{Representative Feynman diagrams for leading-color
$\CA^{(2)}(g,g,g,g,g)$ amplitudes, contributing at order
$N_f^0$, $N_f^1$ and $N_f^2$.}
\label{fig_parents5g}
\end{figure}

\begin{figure}[]
  \begin{center}
    \begin{tikzpicture}[scale=.9]
    % 5 point masters
    \node at (0,0){\includegraphics[scale=0.5]
    {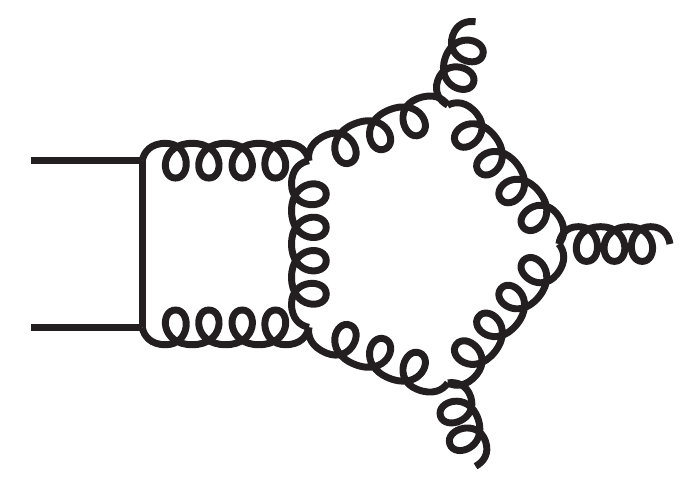}};
    % 5 point masters
    \node at (5,0){\includegraphics[scale=0.5]
    {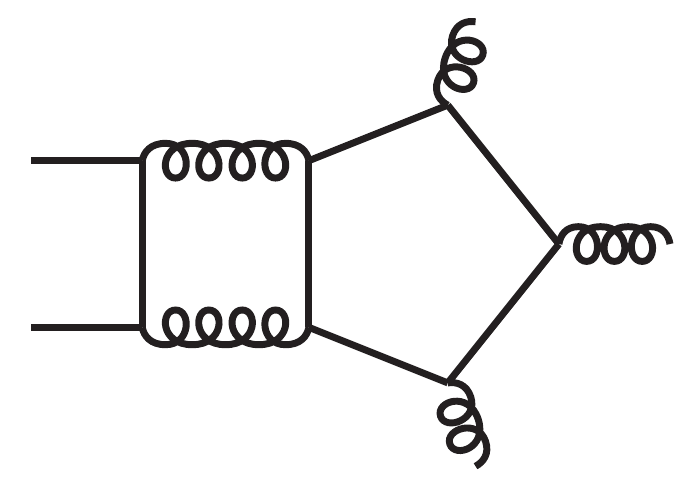}};
    % Level 2 
    \node at (10,0){\includegraphics[scale=0.5]
    {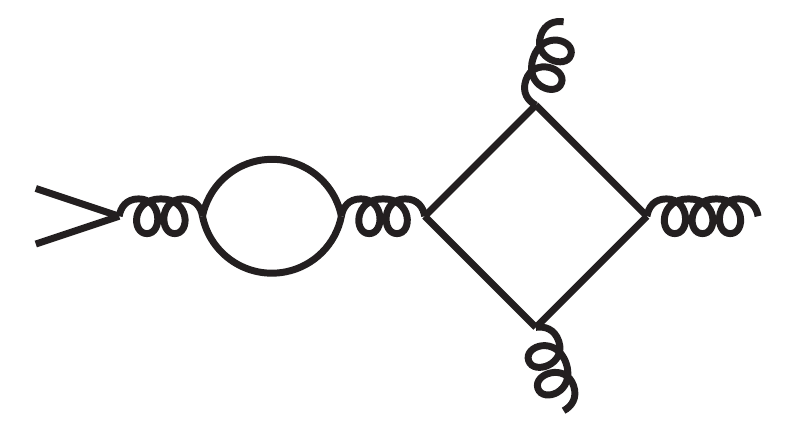}};
\end{tikzpicture}
\end{center} 
\caption{Representative Feynman diagrams for leading-color
$\CA^{(2)}(q,\bar q,g,g,g)$ amplitudes, 
contributing at order
 $N_f^0$, $N_f^1$ and $N_f^2$.}
\label{fig_parents2q3g}
\end{figure}

\begin{figure}[]
  \begin{center}
    \begin{tikzpicture}[scale=.9]
    % 5 point masters
    \node at (0,0){\includegraphics[scale=0.5]
    {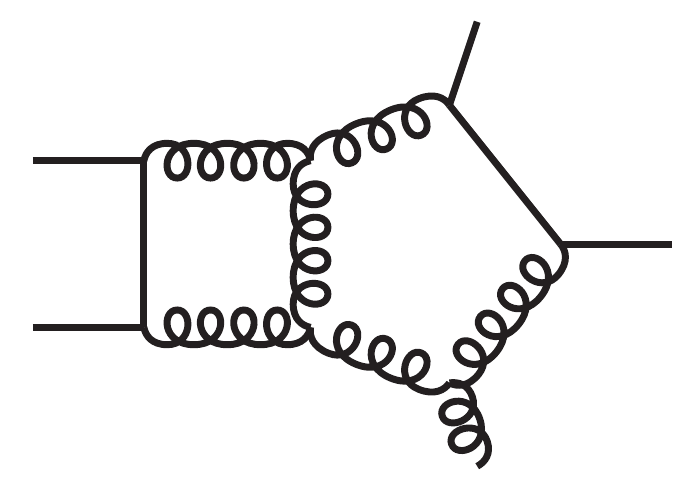}};
    % 5 point masters
    \node at (5,.4){\includegraphics[scale=0.5]
    {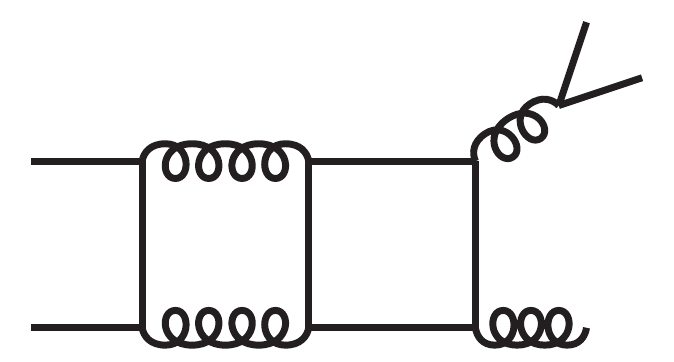}};
    % Level 2 
    \node at (10,.4){\includegraphics[scale=0.5]
    {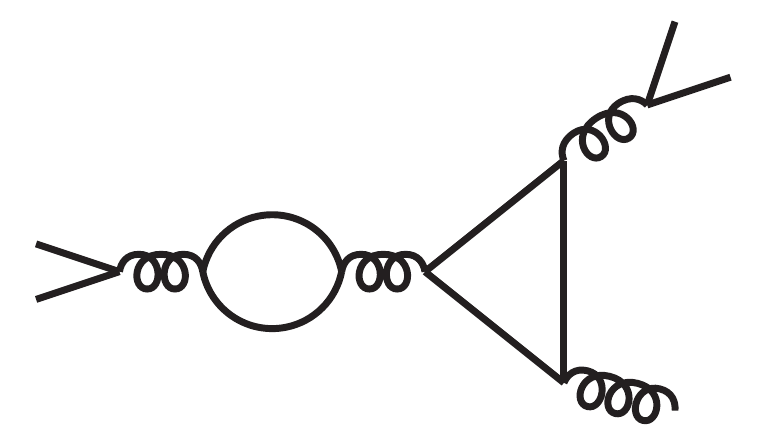}};
\end{tikzpicture}
\end{center} 
\caption{Representative Feynman diagrams for leading-color
$\CA^{(2)}(q,\bar q,Q,\bar Q,g)$ amplitudes, 
contributing at order
$N_f^0$, $N_f^1$ and $N_f^2$.}
\label{fig_parents4q1g}
\end{figure}

The kinematic coefficients of equations 
\eqref{eq:bareAmp4p}--\eqref{eq:ColorDec4Q}, 
denoted by the various $\CA$, are known 
as the leading-color partial amplitudes. They can be
perturbatively expanded up to the two-loop order as
\begin{equation}
    \label{eq:partials} 
    \CA
    = g^{3}_0 \left(
        \CA^{(0)}
      + \frac{\alpha_0}{4\pi}\NC \CA^{(1)}
      + \left(\frac{\alpha_0}{4\pi}\right)^2\NC^2  \CA^{(2)} 
      + \mathcal{O}(\alpha_0^3)
      \right),
\end{equation}
where $\alpha_0=g_0^2/(4\pi)$ is the bare QCD coupling and 
$\CA^{(k)}$ denotes a $k$-loop partial amplitude. 
The partial amplitudes can be further organized in terms of the
number of closed fermion loops, ranging from none up to the loop
order, which each contribute one power of $N_f$. We write 
\begin{align}
  \label{eq:nfdecomposition} 
  \begin{split}
  \CA^{(1)} &= \CA^{(1)[N_f^0]} + 
  \frac{\NF}{\NC}\CA^{(1)[N_f^1]}\,, \\
  \CA^{(2)} &= \CA^{(2)[N_f^0]} +
  \frac{\NF}{\NC}\CA^{(2)[N_f^1]} +
  \left(\frac{\NF}{\NC}\right)^2\CA^{(2)[N_f^2]}\,.
  \end{split} 
\end{align}
In the leading-color approximation, the structure of these
amplitudes simplifies, receiving contributions only from planar
diagrams.
Representative diagrams for each of the five-parton amplitudes 
we consider are given in figs.~\ref{fig_parents5g},
\ref{fig_parents2q3g} and \ref{fig_parents4q1g}.
%%%%%%%%%%%%% FIGURE %%%%%%%%%%%%%%%%%%

%%%%%%%%%%%%% FIGURE %%%%%%%%%%%%%%%%%%

%%%%%%%%%%%%% FIGURE %%%%%%%%%%%%%%%%%%

% !TEX root = ../main.tex

\section{Calculation of Planar Multi-Parton Amplitudes}
\label{sec:Implementation}

In order to compute two-loop four- and five-parton
amplitudes, we apply a variant of the unitarity method
\cite{Bern:1994zx,Bern:1994cg,Bern:1997sc,Britto:2004nc} suitable for
automated numerical computations of multi-loop
amplitudes~\cite{Ita:2015tya,Abreu:2017xsl,Abreu:2017idw}.
The aim of the computation is to
determine the coefficient functions $c_{\Gamma,i}$ and combine them with the master integrals ${\mathcal I}_{\Gamma,i}$ 
in the standard decomposition of the amplitude:
\begin{equation}\label{eq:A}
\CA^{(k)}=\sum_{\Gamma\in \Delta} \sum_{i\,\in\, M_\Gamma} c_{\Gamma,i} \,
{\mathcal I}_{\Gamma,i}\,.
\end{equation}
Here $\Delta$ is the set of all diagrams that specify 
different propagator structures $\Gamma$ in the amplitude.
The index $i$ runs over the set $M_\Gamma$ of master integrals associated with each propagator structure.

In order to determine the coefficient functions $c_{\Gamma,i}$, we promote
eq.~\eqref{eq:A} to the integrand level. The integrand is 
denoted ${\cal A}(\ell_l)$, where $\ell_l$ represents the loop momenta,
and we decompose it as~\cite{Ita:2015tya}
\begin{equation}\label{eq:AL}
{\cal A}^{(k)}(\ell_l)=\,\sum_{\Gamma \in \Delta}\,\,\sum_{i\,\in\, M_\Gamma\cup
S_\Gamma} \frac{ c_{\Gamma,i} \,m_{\Gamma,i}(\ell_l)}{\prod_{j\in P_\Gamma}
\rho_j}\,, 
\end{equation}
where $P_\Gamma$ is the set of propagators in the diagram $\Gamma$, and the
$\rho_j$ denote inverse propagators. 
We extended the sum in eq.~\eqref{eq:A}
to also run over \emph{surface terms} 
contained in the set $S_\Gamma$. These surface terms vanish upon integration but
they are necessary to parametrize the integrand. 
The surface terms are constructed from a complete set of so-called {\it
unitarity-compatible} integration-by-parts identities
\cite{Gluza:2010ws,Schabinger:2011dz,Ita:2015tya,Larsen:2015ped}. 
For all the processes considered in this article
we use the master/surface-term parametrization given in 
ref.~\cite{Abreu:2017hqn}, which only depends
on the kinematics of the processes.
While in amplitudes with fermions additional
Lorentz-symmetry breaking terms may 
appear prior to integration
(see e.g.~\cite{Badger:2017gta,Anger:2018ove} and the
discussion below eq.~\eqref{eq:etaetaP}),
they do not in our
definition of helicity amplitudes in 
eq.~\eqref{eqn:helampl}. 
The cancellation of these
terms will be discussed in \sect{sec:FiniteFields}.

In the numerical unitarity method, the coefficients 
$c_{\Gamma,i}$ in
the ansatz~\eqref{eq:AL} can be determined by
building systems of linear equations through sampling of 
on-shell values of the loop momenta $\ell_l$.
In the
on-shell limit the leading contributions of eq.~\eqref{eq:AL}
factorize,
\begin{eqnarray}\label{eqn:cutEqn}
\sum_{\rm states}\prod_{i\in T_\Gamma} {\cal A}^{\rm tree}_i(\ell_l^\Gamma) =
\sum_{\substack{\Gamma' \ge \Gamma\,,\\ i\,\in\,M_{\Gamma'}\cup S_{\Gamma'}} }
\frac{ c_{\Gamma',i}\,m_{\Gamma',i}(\ell_l^\Gamma)}{\prod_{j\in
    (P_{\Gamma'}\setminus P_\Gamma) } \rho_j(\ell_l^\Gamma)}\,, \label{eq:CE} 
\end{eqnarray}
where we label the set of the tree amplitudes associated to
the vertices in the diagram $\Gamma$ by $T_\Gamma$, 
and the sum on 
the left-hand side represents the sum over all internal states 
on the internal edges of the diagram $\Gamma$.

In eq.~\eqref{eq:CE} the loop momenta $\ell_l^\Gamma$ is such 
that all propagators in $P_\Gamma$ are on-shell, and so in these
limits we also probe diagrams $\Gamma'$ such that
$P_\Gamma\subseteq P_{\Gamma'}$ (a relation that we
denote as $\Gamma' \ge \Gamma$). Beyond one loop there exist 
diagrams in $\Delta$ with doubled propagators. The numerators of
such diagrams correspond to leading and subleading terms in 
their on-shell limits, and for the latter no factorization of 
the integrand into tree amplitudes is known. 
Nevertheless, as shown in ref.~\cite{Abreu:2017idw}, one can
systematically organize the set of cut equations~\eqref{eqn:cutEqn} 
in such a way that all master-integral
coefficients necessary to obtain the full amplitude can be computed.

In the following, we discuss the details of the procedure when 
applied to processes with fermionic degrees of freedom. First, 
in section~\ref{sec:FiniteFields}, we discuss an approach 
that allows the use of finite fields in the presence of
fermions. Next, in section~\ref{sec:BerendsGiele}, we 
discuss the implementation of the products of tree amplitudes 
with fermions.
Finally, in section~\ref{sec:AmplitudeAssembly}, we describe
how these components come together to compute the integrated
amplitude.

\subsection{Finite Fields and Spinors}
\label{sec:FiniteFields}

The extension of unitarity approaches to employ only operations
defined in an algebraic field was proposed 
in ref.~\cite{Peraro:2016wsq}.
A finite-field based calculation allows to compute exact values
for the integral coefficients $c_{\Gamma,i}$ of eq.~\eqref{eq:A}
in a numerical framework.
This idea
was applied recently in \cite{Badger:2017jhb,Abreu:2017hqn} for
pure gluon-scattering amplitudes, and here we discuss our
implementation for amplitude computations with fermions.

From here on we denote by $\mathbb{F}$ an arbitrary 
number field. In practice, we will be interested in
$\mathbb{F}$ being the field of rational numbers $\mathbb{Q}$
or the finite field $\mathbb{Z}_p$
of all integers modulo a prime number $p$.
In general, polynomial equations do not have solutions in 
$\mathbb{F}$. This is at odds with the fact that
in a unitarity-based approach one needs 
to generate loop momenta which satisfy a set of quadratic
conditions corresponding to setting propagators to zero.
In ref.~\cite{Abreu:2017hqn}, this was resolved
by making sure that all scalar products
between the momenta in the problem were $\mathbb{F}$-valued.
In the presence of fermions, the situation becomes more 
complicated due to the extension of the Clifford algebra beyond 
four dimensions. More specifically, terms such as $\ell^\mu
\gamma_{[D_s]\mu}$ exhibit the $(D-4)$-dimensional
components of the loop momenta,
which are in general not 
$\mathbb{F}$-valued for on-shell momenta
(more concretely, if we work on the field of rational numbers
these components are in general irrational),
leading to terms in the sub-currents of the
Berends-Giele recursion that are not $\mathbb{F}$-valued. 
To address this issue,
we start with a parametrization of the on-shell spaces as in 
ref.~\cite{Abreu:2017hqn} but always use normalized basis
vectors. We write the two-loop momenta as
\begin{equation}
    \ell_1 = (\ell_{1,[4]}, \vec{\mu}_1)\,,\quad \quad \quad
    \ell_2 = (\ell_{2,[4]}, \vec{\mu}_2)\,, 
\end{equation}
where we denote their $(D-4)$-dimensional components as 
$\vec{\mu}_1$ and $\vec{\mu}_2$. Next, we choose an orthonormal
basis $\vec{n}_i$ of the $(D-4)$-dimensional space with
$n_1$ in the direction of $\vec{\mu}_1$ and write
\begin{equation}
\vec{\mu}_1 = r_1 \vec{n}_1, \quad  \vec{\mu}_2 = \frac{\mu_{12}}{\mu_{11}} r_1 \vec{n}_1 + r_2 \vec{n}_2
  \quad \mathrm{where} \quad r_1 = \sqrt{\mu_{11}}, \quad r_2 = 
  \sqrt{\mu_{22} - \mu_{12}^2/\mu_{11}},
\end{equation}
with $\mu_{ij}=\vec{\mu}_i\cdot\vec{\mu}_j$.
In a theory containing only vector particles we only ever need 
the values $r_i^2$, which are $\mathbb{F}$-valued both on- and
off-shell \cite{Abreu:2017hqn}. In contrast, in a theory with
fermions, components of Berends-Giele currents will take the
generic form
\begin{equation}
  \label{eq:ExtendedAlgebra}
  a_{00} + a_{10} r_1 + a_{01} r_2 + a_{11} r_1 r_2, 
\end{equation}
which is not $\mathbb{F}$-valued.
In order to nevertheless be able to
work in the field $\mathbb{F}$,
we consider 
the algebra $\mathbb{V}$ over the field $\mathbb{F}$, with 
$\mathbb{V}$ the vector space spanned by the basis 
$\{r_0=1,r_1,r_2,r_1r_2\}$ and equipped with the standard
addition and multiplication.
All components of the Berends-Giele are elements 
in the algebra, and can thus be written as a linear combination
of the $r_i$ with $\mathbb{F}$-valued coefficients. More
concretely, this means we only need to determine the $a_{ij}$ 
in eq.~\eqref{eq:ExtendedAlgebra} which are $\mathbb{F}$-valued
by construction.

An important observation is that, although the coefficients 
$a_{10}$, $a_{01}$ and $a_{11}$ in 
eq.~\eqref{eq:ExtendedAlgebra} are non-zero in
intermediate stages of the calculations, they vanish
for the integrands of helicity amplitudes as defined in 
eq.~\eqref{eq:tensorDecomposition}. This cancellation of the
$r_i$ terms holds in the HV scheme and
is due to the projection onto the invariant tensors 
$v_n$ of eq.~\eqref{eq:tensorDecomposition}, 
which yields polynomials in the Lorentz invariants 
$\mu_{ij}$ at the integrand level. To see this point more
explicitly, consider the
integrand $M_k(\ell_l)$
of an amplitude with an arbitrary
number of quark lines where the subscript $k$ encodes the
dependence on the $(D_s-4)$ spinor 
indices.\footnote{This can be viewed as a generalization of the
decomposition in eq.~\eqref{eq:tensorDecomposition} to the
integrand level.} We can write the integrand in the form
\begin{equation}
  M_k(\ell_l) = \sum_{n,m} f^{\rho_1 \cdots \rho_n, \sigma_1
  \cdots \sigma_m}_{k}
  \left(\prod_{i=1}^n \vec{\mu}_{1 \, \rho_i}\right)
  \left(\prod_{j=1}^m \vec{\mu}_{2 \, \sigma_i}\right),
\end{equation}
with the tensors $f^{\rho_1 \cdots \rho_n, \sigma_1\cdots \sigma_m}_{k}$ implicitly defined. By construction, they
depend on the $(D_s-4)$ components of the loop momenta through
the Lorentz invariant scalar products $\mu_{ij}$. The
$(D_s-4)$ Lorentz indices we write explicitly can only be 
carried in $f^{\rho_1 \cdots \rho_n, \sigma_1\cdots \sigma_m}_{k}$ by $(D_s-4)$-dimensional $\gamma$-matrices or
metric tensors.
In our definition \eqref{eq:tensorDecomposition} of helicity 
amplitudes, the $(D_s-4)$-dimensional spinor indices are to be
contracted with invariant tensors, leading to traces of
$\gamma_{[D_s-4]}$ matrices which
can be expressed in terms of metric tensors. Consequently, the
Lorentz indices in a contraction of $f^{\rho_1 \cdots \rho_n, \sigma_1\cdots \sigma_m}_{k}$ with an invariant tensor
are carried by metric tensors only.
Hence, after contraction with invariant tensors, 
the integrand only depends on $\mu_{ij}$.
In contrast, evaluating amplitudes that introduce a reference
axis in the $(D_s-4)$-dimensional space
would lead in general to a dependence on the components of the
$\vec\mu_i$ and thus on the $r_i$ terms. 
This is the case for instance when considering
gluon-polarization components in the $(D_s-4)$ dimensions 
(as required in the CDR scheme) or generic values of the 
$(D_s-4)$-dimensional spinors $\eta_i$, 
as written explicitly in eq.~\eqref{eq:etaetaP}.

We finish with a comment that is not related to the use of 
finite fields but follows from the discussion above.
Since our representation of fermion amplitudes is manifestly
Lorentz invariant in $(D_s-4)$ dimensions prior to loop
integration, the integrands of fermion amplitudes can be
decomposed in terms of the same set of master integrands and
surface terms as those used for amplitudes with gluons only
\cite{Abreu:2017xsl,Abreu:2017hqn}.

\subsection{Tree Amplitudes}
\label{sec:BerendsGiele}
In order to numerically calculate the necessary products of tree
amplitudes used in the cut equations \eqref{eqn:cutEqn}, we
implement a Berends-Giele recursion~\cite{Berends:1987me}. The
presence of the fermionic degrees of freedom means that we
require concrete representations of the Clifford algebra in 
$D_s$ dimensions, where $D_s$ is even. It can be
shown that integrands of the HV amplitudes defined in eqs.~\eqref{eq:qqProj} and \eqref{eq:AmplitudeDefinition}
depend at most quadratically on the
parameter $D_s$. As we must also take $D_s \ge 6$, 
we implement the recursion for three values of $D_s$,
specifically $6$, $8$ and $10$. Explicit constructions can be
found (for example) in 
\cite{Kreuzer:susylectures,Collins:1984xc} or obtained using
the factorized definition in~\eqref{eqn:cliffordrecursion}.
Importantly, to obtain manifestly real representations of the Clifford algebra, we 
continue components of momenta to imaginary values keeping
kinematic invariants real valued.  
For gluon amplitudes, the analytic
continuation can be equivalently interpreted as changing the
metric signature to
$g'_{[D_s]}={\rm diag}\{ +1 , -1 , +1, ... , -1 \}$.
As far as spinor representations are concerned,
the two perspective are not equivalent as the latter would also
alter the inner product of the spinors.
Effectively we work in the alternating signature while
maintaining the conjugation operation for spinors as 
defined in Minkowski signature. 

In order to implement the prescription of 
eqs.~\eqref{eq:qqProj} and \eqref{eq:AmplitudeDefinition}
for computing amplitudes with external fermions, we
first construct four-dimensional states with a specific 
helicity from Weyl spinors 
using the conventions of ref.~\cite{Maitre:2007jq}.
To handle the $(D_s-4)$-dimensional
Clifford algebra we work with a canonical basis for the
associated spinors $\eta^i_\kappa=\delta^i_\kappa$
and fix $\bar \eta_i^\kappa$ to be its dual as in 
eq.~\eqref{eqn:qspinors}.
Through eq.~\eqref{eq:fermionicStates} we
then construct the full set of $D_s$-dimensional states
associated with a given four-dimensional state.
The projections in eqs.~\eqref{eq:qqProj} and \eqref{eq:AmplitudeDefinition}
then amount to the evaluation of (normalized) traces over the $(D_s-4)$-dimensional indices.

We close with two technical remarks.
First, within our implementation all internal Lorentz indices
are taken to be $D_s$-dimensional despite the HV prescription
that this should only be the case for one-particle-irreducible diagrams.
This is allowed because the difference between this prescription
and the HV prescription does not contribute to the helicity
amplitudes as defined in 
eqs.~\eqref{eq:qqProj} and \eqref{eq:AmplitudeDefinition}.
Second, with an appropriate normalization of the spinor states
and their conjugates, the components of the
spinors in internal state-sums also take the form of
eq.~\eqref{eq:ExtendedAlgebra}, so no special treatment is
needed in finite-field computations.

\subsection{Amplitude Evaluation}
\label{sec:AmplitudeAssembly}
We start by constructing the set $\Delta$ of all propagator
structures which are associated to a given amplitude in the
decomposition of eq.~\eqref{eq:AL}. 
For this task we produce all 
cut diagrams in the full-color process employing 
QGRAF \cite{Nogueira:1991ex}, followed by a color decomposition
performed in \Mathematica{} according to 
ref.~\cite{Ochirov:2016ewn}. In the latter step, tree-level
decompositions for processes involving several fermion
lines are necessary and we perform them following 
ref.~\cite{FermionColour}. 
We then take the leading-color limit and extract a hierarchically-organized set
of propagator structures associated to the color-ordered amplitudes in
eqs.~\eqref{eq:bareAmp4p}--\eqref{eq:ColorDec4Q}.
This decomposition is then processed by a \texttt{C++} code.
The master/surface-term decomposition which we employ is 
the same as the one used in 
refs.~\cite{Abreu:2017xsl,Abreu:2017hqn}.
It was constructed using the computational algebraic
geometry package {\sc SINGULAR} \cite{DGPS} to solve
syzygy equations that allow to obtain a set of
unitarity-compatible surface terms.

Solving the multiple systems of linear equations associated to all cut equations~\eqref{eqn:cutEqn}
is achieved through PLU factorization and back substitution. To reconstruct the
dependence of the master-integral coefficients on the dimensional regulators $D$
and $D_s$ we sample over enough values to resolve their rational or polynomial
dependence, respectively.
In a generalization of ref.~\cite{Ellis:2008ir}, the
quadratic $D_s$ dependence is reconstructed from the evaluations
at $D_s=6$, $8$ and $10$.
Explicit $D$ dependence on the integral coefficients is 
induced by the $D$-dependent surface terms. We sample multiple
values of $D$ randomly to extract the rational dependence of all
master coefficients by using Thiele's 
formula~\cite{Peraro:2016wsq, abramowitz1964handbook}.

The above approach to obtaining the coefficients in the
decomposition of the amplitude in eq.~\eqref{eq:A} is
implemented in a numerical framework which allows two
independent computations. The first involves
evaluation over the finite fields provided by 
Givaro~\cite{Givaro}.
We use cardinalities of order $2^{30}$ and, to improve on
the multiplication speed, we implement Barrett
reduction~\cite{Barrett1987,HoevenLQ14}.
For a given rational kinematic point, 
we perform the computation in a sufficient 
(phase-space-point dependent) number of finite fields to apply a
rational-reconstruction algorithm after using the Chinese
Remainder theorem.
The second mode of operation carries out the evaluations 
in high-precision floating-point arithmetic.
In this case we do neither employ the technology to control
algebraic terms described in section~\ref{sec:FiniteFields}
nor do we
use the refined computational setup based on real-valued operations.

The coefficients are then combined with master integrals as
in eq.~\eqref{eq:A} to give the integrated amplitudes. For
four-parton amplitudes we use the same implementation of
the integrals as the one used in ref.~\cite{Abreu:2017xsl} 
and for five-parton amplitudes we use the same as in 
ref.~\cite{Abreu:2017hqn}. 
In the former case, we used our own calculation of a set of 
master integrals. In the latter case, we used the
integrals of ref.~\cite{Papadopoulos:2015jft} for the five-point
master integrals, the integrals of ref.~\cite{Gehrmann:2000zt}
for the lower point integrals, and our own calculation of the 
one-loop-factorizable integrals. In all cases, the 
polylogarithms in the $\epsilon$-expansion of the master
integrals are evaluated with GiNaC~\cite{Vollinga:2004sn}, which
can be tuned to the desired precision.

% !TEX root = ../main.tex

\section{Numerical Results for Helicity Amplitudes}
\label{sec:Results}

In this section we present numerical values for 
leading-color two-loop multi-parton helicity amplitudes.
We first present our results for four-parton amplitudes. These
are known in analytic 
form
\cite{Bern:2002tk,Bern:2003ck,DeFreitas:2004kmi,Glover:2004si}
and we use  them as a validation of our approach.
Then we present our new computation of five-parton helicity
amplitudes. We include all $N_f$ corrections
corresponding to closed massless-quark loops.

For each different choice of external partons we consider,
we will show tables of numerical results for a full set of
independent helicity assignments
corresponding to a single partial 
amplitude in the color decompositions of 
eqs.~\eqref{eq:bareAmp4p}--\eqref{eq:ColorDec4Q}.
Furthermore,
we present results only for distinct flavor
configurations: as discussed in section~\ref{sec:Methods},
see eq.~\eqref{eq:LikeInterference},
results for finite remainders of amplitudes with identical 
quarks can be obtained by antisymmetrizing on the flavor
assignments. In appendix~\ref{sec:IR} we give all the 
ingredients required for computing these remainders, in 
particular results for one-loop amplitudes expanded through 
order $\epsilon^2$.

\subsection{Four-parton Amplitudes}
We evaluate the four-gluon, two-quark two-gluon and four-quark
amplitudes at the phase-space point%
\footnote{Units of energy are chosen arbitrarily.
  The amplitudes presented in the tables
  \ref{tab:results4parton}, \ref{tab:results5parton}, 
  \ref{tab:results4parton1L} and \ref{tab:results5parton1L}
  are normalized to be dimensionless.
}
\begin{equation}
  \begin{aligned}[c]
    p_1 &= \vphantom{\frac{1}{48}}\left(1,1,-i,1\right), \\
    p_2 &= -\frac{1}{16}\left(3,0,0,-3\right), \\
  \end{aligned}
  \qquad
  \begin{aligned}[c]
    p_3 &= \frac{1}{48}\left(25,-51,45\,i,7\right),\\
    p_4 &= -\frac{1}{48}\left(64,-3,-3\,i, 64\right), \\
  \end{aligned}
  \label{eq:EvalPoint4}
\end{equation}
with corresponding invariants $s_{12}=-3/4$ and $s_{23}=-1/4$ 
where $s_{ij}=(p_i+p_j)^2$.
We set the regularization scale $\mu$ to 1 and the normalization
of the results is fixed by the expansion in 
\eqns{eq:partials}{eq:nfdecomposition}. All results are presented
in the HV scheme.
\begin{table}[h]
  \begin{adjustbox}{width=1\textwidth}
    \centering
    \begin{tabular}{cccccc}
      \toprule
      $\CA^{(2)[N_f^0]}/\CA^{\rm (norm)}$	  &   $\epsilon^{-4}$   &   $\epsilon^{-3}$   &   $\epsilon^{-2}$   &   $\epsilon^{-1}$   &   $\epsilon^{0}$   \\
      \midrule
      $( 1_g^+, 2_g^+, 3_g^+, 4_g^+ )$ & 0 & 0 & $-4.000000000$ & $-23.74072126$ & $-63.52221777$ \\
      $( 1_g^-, 2_g^+, 3_g^+, 4_g^+ )$ & 0 & 0 & $-4.000000000$ & $-35.31127327$ & $-133.5083818$ \\
      $( 1_g^-, 2_g^-, 3_g^+, 4_g^+ )$ & $8.000000000$ & $55.65274878$ & $164.6421815$ & $222.3267401$ & $-8.390444844$ \\
      $( 1_g^-, 2_g^+, 3_g^-, 4_g^+ )$ & $8.000000000$ & $55.65274878$ & $176.0091465$ & $332.2956004$ & $486.5023259$ \\
      \midrule
      $( 1_q^+, 2_{\bar q}^-, 3_g^+, 4_g^+ )$ & 0 & 0 & $-3.000000000$ & $-24.41444952$ & $-74.97642231$ \\
      $( 1_q^+, 2_{\bar q}^-, 3_g^+, 4_g^- )$ & $4.500000000$ & $28.51508962$ & $73.34964082$ & $75.65107559$ & $-9.311163231$ \\
      $( 1_q^+, 2_{\bar q}^-, 3_g^-, 4_g^+ )$ & $4.500000000$ & $28.51508962$ & $64.00475414$ & $-13.64171730$ & $-376.4555455$ \\
      \midrule
      $( 1_q^+, 2_{\bar q}^-, 3_Q^+, 4_{\bar Q}^- )$ & $2.000000000$ & $10.19374511$ & $8.003461515$ & $-55.57160018$ & $-92.52942183$ \\
      $( 1_q^+, 2_{\bar q}^-, 3_Q^-, 4_{\bar Q}^+ )$ & $2.000000000$ & $10.19374511$ & $-4.028725695$ & $-134.3060579$ & $-234.1564069$ \\
      \toprule
      $\CA^{(2)[N_f^1]}/\CA^{\rm (norm)}$	  &   $\epsilon^{-4}$   &   $\epsilon^{-3}$   &   $\epsilon^{-2}$   &   $\epsilon^{-1}$   &   $\epsilon^{0}$   \\
      \midrule
      $( 1_g^+, 2_g^+, 3_g^+, 4_g^+ )$ & 0 & 0 & $4.000000000$ & $27.74072126$ & $86.81849458$ \\
      $( 1_g^-, 2_g^+, 3_g^+, 4_g^+ )$ & 0 & 0 & $4.000000000$ & $39.31127327$ & $172.4199379$ \\
      $( 1_g^-, 2_g^-, 3_g^+, 4_g^+ )$ & 0 & $-2.000000000$ & $-15.96133691$ & $-59.69423578$ & $-141.8161833$ \\
      $( 1_g^-, 2_g^+, 3_g^-, 4_g^+ )$ & 0 & $-2.000000000$ & $-18.16301631$ & $-81.04594245$ & $-230.6319267$ \\
      \midrule
      $( 1_q^+, 2_{\bar q}^-, 3_g^+, 4_g^+ )$ & 0 & 0 & $0.5454545455$ & $3.784151849$ & $3.326492162$ \\
      $( 1_q^+, 2_{\bar q}^-, 3_g^+, 4_g^- )$ & 0 & $0.5000000000$ & $4.307232180$ & $15.70646205$ & $21.70488360$ \\
      $( 1_q^+, 2_{\bar q}^-, 3_g^-, 4_g^+ )$ & 0 & $0.5000000000$ & $4.307232180$ & $13.62982056$ & $-12.51632628$ \\
      \midrule
      $( 1_q^+, 2_{\bar q}^-, 3_Q^+, 4_{\bar Q}^- )$ & 0 & $1.666666667$ & $10.55774898$ & $23.90612711$ & $-30.33285238$ \\
      $( 1_q^+, 2_{\bar q}^-, 3_Q^-, 4_{\bar Q}^+ )$ & 0 & $1.666666667$ & $10.55774898$ & $15.88466897$ & $-106.4874291$ \\
      \toprule
      $\CA^{(2)[N_f^2]}/\CA^{\rm (norm)}$	  &   $\epsilon^{-4}$   &   $\epsilon^{-3}$   &   $\epsilon^{-2}$   &   $\epsilon^{-1}$   &   $\epsilon^{0}$   \\
      \midrule
      $( 1_g^+, 2_g^+, 3_g^+, 4_g^+ )$ & 0 & 0 & 0 & 0 & $1.444444444$ \\
      $( 1_g^-, 2_g^+, 3_g^+, 4_g^+ )$ & 0 & 0 & 0 & 0 & $0$ \\
      $( 1_g^-, 2_g^-, 3_g^+, 4_g^+ )$ & 0 & 0 & 0 & 0 & $0.03086419753$ \\
      $( 1_g^-, 2_g^+, 3_g^-, 4_g^+ )$ & 0 & 0 & 0 & 0 & $0$ \\
      \midrule
      $( 1_q^+, 2_{\bar q}^-, 3_g^+, 4_g^+ )$ & 0 & 0 & 0 & $0.1212121212$ & $1.189856320$ \\
      $( 1_q^+, 2_{\bar q}^-, 3_g^+, 4_g^- )$ & 0 & 0 & 0 & 0 & $0$ \\
      $( 1_q^+, 2_{\bar q}^-, 3_g^-, 4_g^+ )$ & 0 & 0 & 0 & 0 & $0$ \\
      \midrule
      $( 1_q^+, 2_{\bar q}^-, 3_Q^+, 4_{\bar Q}^- )$ & 0 & 0 & $0.4444444444$ & $3.473917619$ & $14.37639897$ \\
      $( 1_q^+, 2_{\bar q}^-, 3_Q^-, 4_{\bar Q}^+ )$ & 0 & 0 & $0.4444444444$ & $3.473917619$ & $14.37639897$ \\
      \bottomrule
    \end{tabular}
  \end{adjustbox}
  \caption{The bare two-loop four-parton helicity amplitudes evaluated
    at the phase space point in eq.~\eqref{eq:EvalPoint4}. We set the
    normalization factor $\CA^{\rm (norm)}$ to $\CA^{(1)[N_f^0]}(\epsilon=0)$ for the
    amplitudes with vanishing trees, and to $\CA^{(0)}$ otherwise.} 
  \label{tab:results4parton}
\end{table}

In table~\ref{tab:results4parton} we show numerical results for
the bare two-loop four-parton helicity amplitudes.
In order to expose the pole structure of the amplitudes (see
appendix~\ref{sec:IR}) we normalize them to
the corresponding tree-level amplitude if
it is nonvanishing, or to the corresponding $\CA^{(1)[N_f^0]}(\epsilon=0)$ amplitude otherwise.
The results have been obtained with exact values for the 
integral coefficients and with the master integrals 
evaluated to a precision that allows to show 10
significant digits.

We have validated our results by carrying out a set of
checks. We verified that they satisfy the
expected infrared pole structure \cite{Catani:1998bh}. We summarize the
relevant formulae for this check in appendix~\ref{sec:IR}.
Furthermore, we have carried out a systematic
validation of the $\epsilon^0$ contributions of all of 
our results against their known analytic
expressions from refs.~\cite{Bern:2002tk,Bern:2003ck,
DeFreitas:2004kmi,Glover:2004si}. 
For the four-gluon amplitudes we have compared directly the
$\epsilon^0$ pieces of our results with the analytic expressions
of~\cite{Bern:2002tk}. For the
two-quark two-gluon and four-quark amplitudes we have used the
one-loop results given in appendix~\ref{sec:oneloopvalues}
to compute the corresponding finite remainders $\CF^{(2)}$ as
defined in eq.~\eqref{eq:remainderDef}. 
After accounting for the different choices of 
normalization for the $\mathbf{H}_{[n]}(\epsilon)$ operators 
(see appendix~\ref{sec:IR})
made in refs.~\cite{Bern:2003ck} and \cite{Glover:2004si},
we have found perfect agreement.

\subsection{Five-parton Amplitudes}
\label{sec:fivepartonRes}
We present results for the five-parton amplitudes evaluated at
the phase-space point
\begin{equation}
  \begin{aligned}[c]
    p_1 &= \left( \frac{1}{2}, \frac{45}{272}, \frac{45 i}{272}, \frac{1}{2} \right), \\
    p_2 &= \left( -\frac{1}{2}, 0, 0, \frac{1}{2} \right), \\
    p_3 &= \left( \frac{21}{26}, -\frac{21}{26}, -\frac{5 i}{26}, -\frac{5}{26} \right),
  \end{aligned}
  \qquad
  \begin{aligned}[c]
    p_4 &= \left( -\frac{1169}{2652}, \frac{2165}{10608}, -\frac{13459 i}{38896}, -\frac{5075}{9724} \right),\\[3ex]
    p_5 &= \left( -\frac{973}{2652}, \frac{581}{1326}, \frac{1813 i}{4862}, -\frac{2779}{9724} \right),
  \end{aligned}
  \label{eq:EvalPoint5}
\end{equation}
with corresponding invariants
\begin{align}\begin{split}
  s_{12} =& -1, \quad s_{23} = -8/13, \quad s_{34} = -1094/2431,\\
  &s_{45} = -7/17, \quad s_{51} = -749/7293\ .
\label{eq:Invs5pt}
\end{split}\end{align}
We set the regularization scale~$\mu$ to 1 and the normalization
of the results is fixed by the expansion in 
\eqns{eq:partials}{eq:nfdecomposition}. All results have been computed in the 
HV scheme. Given that our integral coefficients are computed as 
exact rational numbers, the final precision of our results is 
determined by how many digits we require from 
GiNaC~\cite{Vollinga:2004sn}
in the evaluation of the polylogarithms in the master integrals
we use \cite{Papadopoulos:2015jft,Gehrmann:2000zt}. In
table~\ref{tab:results5parton} we present results with 10
significant digits.
\begin{table}[!htbp]
  \centering
  \begin{adjustbox}{width=1\textwidth}
    \begin{tabular}{cccccc}
      \toprule
      $\CA^{(2)[N_f^0]}/\CA^{\rm (norm)}$	  &   $\epsilon^{-4}$   &   $\epsilon^{-3}$   &   $\epsilon^{-2}$   &   $\epsilon^{-1}$   &   $\epsilon^{0}$   \\
      \midrule
      $( 1_g^+, 2_g^+, 3_g^+, 4_g^+, 5_g^+ )$ & 0 & 0 & $-5.000000000$ & $-29.38541207$ & $-62.68413553$ \\
      $( 1_g^-, 2_g^+, 3_g^+, 4_g^+, 5_g^+ )$ & 0 & 0 & $-5.000000000$ & $-42.33840431$ & $-159.9778589$ \\
      $( 1_g^-, 2_g^-, 3_g^+, 4_g^+, 5_g^+ )$ & $12.50000000$ & $84.83123596$ & $243.4660216$ & $301.9565843$ & $-152.0528809$ \\
      $( 1_g^-, 2_g^+, 3_g^-, 4_g^+, 5_g^+ )$ & $12.50000000$ & $84.83123596$ & $269.4635002$ & $551.6251881$ & $984.0882231$ \\
      \midrule
      $( 1_q^+, 2_{\bar q}^-, 3_g^+, 4_g^+, 5_g^+ )$ & 0 & 0 & $-4.000000000$ & $-33.66432052$ & $-117.5792214$ \\
      $( 1_q^+, 2_{\bar q}^-, 3_g^+, 4_g^+, 5_g^- )$ & $8.000000000$ & $51.38308777$ & $127.3357346$ & $55.24748112$ & $-511.9128286$ \\
      $( 1_q^+, 2_{\bar q}^-, 3_g^+, 4_g^-, 5_g^+ )$ & $8.000000000$ & $51.38308777$ & $137.2047686$ & $143.1002284$ & $-154.2224796$ \\
      $( 1_q^+, 2_{\bar q}^-, 3_g^-, 4_g^+, 5_g^+ )$ & $8.000000000$ & $51.38308777$ & $133.2453937$ & $110.9941406$ & $-263.9507190$ \\
      \midrule
      $( 1_q^+, 2_{\bar q}^-, 3_Q^+, 4_{\bar Q}^-, 5_g^+ )$ & $4.500000000$ & $23.78050411$ & $33.01035431$ & $-76.65528489$ & $-305.7123751$ \\
      $( 1_q^+, 2_{\bar q}^-, 3_Q^-, 4_{\bar Q}^+, 5_g^+ )$ & $4.500000000$ & $23.78050411$ & $25.33119767$ & $-122.8050519$ & $-400.0885233$ \\
      $( 1_q^+, 2_{\bar q}^-, 3_Q^+, 4_{\bar Q}^-, 5_g^- )$ & $4.500000000$ & $23.78050411$ & $25.00917906$ & $16.91995611$ & $579.1225796$ \\
      $( 1_q^+, 2_{\bar q}^-, 3_Q^-, 4_{\bar Q}^+, 5_g^- )$ & $4.500000000$ & $23.78050411$ & $-1009.208812$ & $-4797.768367$ & $4827.790534$ \\
      \midrule
      $\CA^{(2)[N_f^1]}/\CA^{\rm (norm)}$	  &   $\epsilon^{-4}$   &   $\epsilon^{-3}$   &   $\epsilon^{-2}$   &   $\epsilon^{-1}$   &   $\epsilon^{0}$   \\
      \midrule
      $( 1_g^+, 2_g^+, 3_g^+, 4_g^+, 5_g^+ )$ & 0 & 0 & $5.000000000$ & $34.38541207$ & $78.06348509$ \\
      $( 1_g^-, 2_g^+, 3_g^+, 4_g^+, 5_g^+ )$ & 0 & 0 & $5.000000000$ & $47.33840431$ & $206.9626532$ \\
      $( 1_g^-, 2_g^-, 3_g^+, 4_g^+, 5_g^+ )$ & 0 & $-2.500000000$ & $-15.82327813$ & $-36.65791641$ & $-15.54781774$ \\
      $( 1_g^-, 2_g^+, 3_g^-, 4_g^+, 5_g^+ )$ & 0 & $-2.500000000$ & $-20.72836557$ & $-83.86917083$ & $-215.3966037$ \\
      \midrule
      $( 1_q^+, 2_{\bar q}^-, 3_g^+, 4_g^+, 5_g^+ )$ & 0 & 0 & $1.416882412$ & $11.98234731$ & $38.78056708$ \\
      $( 1_q^+, 2_{\bar q}^-, 3_g^+, 4_g^+, 5_g^- )$ & 0 & $0.6666666667$ & $7.912904946$ & $38.94492002$ & $78.45710970$ \\
      $( 1_q^+, 2_{\bar q}^-, 3_g^+, 4_g^-, 5_g^+ )$ & 0 & $0.6666666667$ & $5.701796856$ & $20.47669656$ & $20.24036826$ \\
      $( 1_q^+, 2_{\bar q}^-, 3_g^-, 4_g^+, 5_g^+ )$ & 0 & $0.6666666667$ & $5.878666845$ & $21.43074531$ & $17.31964894$ \\
      \midrule
      $( 1_q^+, 2_{\bar q}^-, 3_Q^+, 4_{\bar Q}^-, 5_g^+ )$ & 0 & $2.500000000$ & $17.25407596$ & $48.27686582$ & $11.71960460$ \\
      $( 1_q^+, 2_{\bar q}^-, 3_Q^-, 4_{\bar Q}^+, 5_g^+ )$ & 0 & $2.500000000$ & $17.27259645$ & $44.99884204$ & $-15.14666233$ \\
      $( 1_q^+, 2_{\bar q}^-, 3_Q^+, 4_{\bar Q}^-, 5_g^- )$ & 0 & $2.500000000$ & $3.980556493$ & $-29.18374008$ & $-149.0347042$ \\
      $( 1_q^+, 2_{\bar q}^-, 3_Q^-, 4_{\bar Q}^+, 5_g^- )$ & 0 & $2.500000000$ & $180.9505853$ & $624.1255757$ & $-2759.824817$ \\
      \midrule
      $\CA^{(2)[N_f^2]}/\CA^{\rm (norm)}$	  &   $\epsilon^{-4}$   &   $\epsilon^{-3}$   &   $\epsilon^{-2}$   &   $\epsilon^{-1}$   &   $\epsilon^{0}$   \\
      \midrule
      $( 1_g^+, 2_g^+, 3_g^+, 4_g^+, 5_g^+ )$ & 0 & 0 & 0 & 0 & $13.52483164$ \\
      $( 1_g^-, 2_g^+, 3_g^+, 4_g^+, 5_g^+ )$ & 0 & 0 & 0 & 0 & $0.08295433103$ \\
      $( 1_g^-, 2_g^-, 3_g^+, 4_g^+, 5_g^+ )$ & 0 & 0 & 0 & 0 & $0.2400910586$ \\
      $( 1_g^-, 2_g^+, 3_g^-, 4_g^+, 5_g^+ )$ & 0 & 0 & 0 & 0 & $0.008096515560$ \\
      \midrule
      $( 1_q^+, 2_{\bar q}^-, 3_g^+, 4_g^+, 5_g^+ )$ & 0 & 0 & 0 & $0.2361470687$ & $2.541010053$ \\
      $( 1_q^+, 2_{\bar q}^-, 3_g^+, 4_g^+, 5_g^- )$ & 0 & 0 & 0 & $0.3690523831$ & $3.782474720$ \\
      $( 1_q^+, 2_{\bar q}^-, 3_g^+, 4_g^-, 5_g^+ )$ & 0 & 0 & 0 & $0.0005343680110$ & $0.004830824685$ \\
      $( 1_q^+, 2_{\bar q}^-, 3_g^-, 4_g^+, 5_g^+ )$ & 0 & 0 & 0 & $0.03001269961$ & $0.3139119453$ \\
      \midrule
      $( 1_q^+, 2_{\bar q}^-, 3_Q^+, 4_{\bar Q}^-, 5_g^+ )$ & 0 & 0 & $0.4444444444$ & $3.910872659$ & $18.01752271$ \\
      $( 1_q^+, 2_{\bar q}^-, 3_Q^-, 4_{\bar Q}^+, 5_g^+ )$ & 0 & 0 & $0.4444444444$ & $3.919103985$ & $18.09637714$ \\
      $( 1_q^+, 2_{\bar q}^-, 3_Q^+, 4_{\bar Q}^-, 5_g^- )$ & 0 & 0 & $0.4444444444$ & $-1.988469328$ & $-28.36258323$ \\
      $( 1_q^+, 2_{\bar q}^-, 3_Q^-, 4_{\bar Q}^+, 5_g^- )$ & 0 & 0 & $0.4444444444$ & $76.66487683$ & $646.7253090$ \\
      \bottomrule
    \end{tabular}
  \end{adjustbox}
  \caption{The bare two-loop five-parton helicity amplitudes evaluated
    at the phase space point in eq.~\eqref{eq:EvalPoint5}. We set the
    normalization factor $\CA^{\rm (norm)}$ to $\CA^{(1)[N_f^0]}(\epsilon=0)$ for the
    amplitudes with vanishing trees, and to $\CA^{(0)}$ otherwise.} 
  \label{tab:results5parton}
\end{table}

All results in table~\ref{tab:results5parton} have been 
checked to satisfy the pole structure of two-loop
amplitudes~\cite{Catani:1998bh}. 
The one-loop amplitudes required for these checks
have been obtained from our own setup,
and cross-checked up to order $\epsilon^0$ 
with \BlackHat{}~\cite{Berger:2008sj}. We present their 
numerical values in appendix \ref{sec:IR}.
The $N_f^0$ piece of the all-plus five-gluon amplitude have been
checked to reproduce the analytic result
of \cite{Gehrmann:2015bfy}, and for the other helicity 
configurations we have validated the results of
\cite{Badger:2017jhb} with our implementation.
We also find agreement with the numerical results of the $N_f^0$ terms
of the two-quark three-gluon and four-quark one-gluon two-loop
amplitudes which have been presented in the revised version of
ref.~\cite{Badger:2018gip}.
Finally, we also cross-checked the pole structures of other 
helicity configurations, not explicitly shown. As our setup is a
numerical one, this amounts to internal consistency checks of 
our computational framework.

% !TEX root = ../main.tex

\section{Conclusion}
\label{sec:Conclusion}

We have presented the calculation of the planar two-loop 
four- and five-parton helicity
amplitudes, extending the numerical variant of the
two-loop unitarity method already used in 
\cite{Abreu:2017xsl,Abreu:2017hqn} to amplitudes with fermions. 
Our results include all corrections associated with closed
massless fermions loops.
Numerical results for some of the amplitudes we have computed
have been presented recently \cite{Badger:2018gip}. 
Given our results, the complete 
set of two-loop amplitudes required for a NNLO QCD
calculation of three-jet production at hadron colliders in the 
leading-color approximation are now available. 

We first described a formalism for computing multi-loop helicity
amplitudes with external fermions in dimensional
regularization, consistent with the approaches
of refs.~\cite{Glover:2003cm,Glover:2004si,Anger:2018ove}.
This was achieved by embedding the four-dimensional
external fermionic states in $D_s$ dimensions and preserving the
invariance of the amplitude under Lorentz transformations in the
$(D_s-4)$-dimensional space.
Within this formalism, we precisely stated our definition of
helicity amplitudes and devised a numerical method to compute 
parton scattering amplitudes in the HV scheme.
After interference with the Born amplitudes, changing
to other regularization schemes can be achieved by known
transition rules~\cite{Broggio:2015dga}.

Our computational approach relies on a
parametrization of the two-loop four- and five-point massless
integrand in terms of master integrands
and surface terms.
With our definition of helicity amplitudes, we can reuse the 
same parametrization already used for four- and five-point gluon
amplitudes independently of the type of partons.
We extended the finite-field implementation 
of ref.~\cite{Abreu:2017hqn} to fermion amplitudes,
allowing us to compute exact master-integral coefficients for 
all partonic subprocesses at rational phase-space
points.
The computations were also performed in an alternative
setup using floating-point arithmetic and we find agreement
between the two variants of our numerical method.

We present reference values for helicity 
amplitudes. These are obtained by combining the master-integral
coefficients we compute with the corresponding master integrals,
in particular using the recently obtained analytic expressions
for five-point integrals~\cite{Papadopoulos:2015jft,
Gehrmann:2018yef}. 
We have validated our results in a number of ways: we reproduce
the results for two-loop four-parton helicity amplitudes 
computed from their known analytic 
expressions~\cite{Bern:2002tk,Bern:2003ck,Glover:2004si}, we
find the correct infrared structure of each amplitude and we
validate the finite pieces of recently published five-parton
results~\cite{Badger:2013gxa,
Badger:2015lda,Gehrmann:2015bfy, Dunbar:2016aux, Badger:2017jhb,
Abreu:2017hqn,Badger:2018gip} as detailed in 
section~\ref{sec:fivepartonRes}.

The techniques developed in this paper show the potential for
the automation of two-loop multi-particle amplitude calculations
in the Standard Model. 
Our numerical approach is relatively insensitive to the addition
of scales. Having already implemented vector and spinor
fields, we are now ready to explore processes of
phenomenological relevance that include jets, (massive) gauge
bosons and leptons in the final state. 
While techniques for computing two-loop master integrals 
progress and new methods appear for handling infrared divergent
terms in real-real and real-virtual contributions, we expect to
provide a program that can deliver one- and two-loop
matrix elements necessary for computing precise QCD predictions
for the LHC.

\acknowledgments

We thank J.~Dormans and M.~Zeng for many useful discussions.
We thank Z.~Bern and L.J.~Dixon for providing analytic 
expressions from references~\cite{Bern:2002tk,Bern:2003ck},
which we employed to validate our results for the two-loop
four-gluon and two-quark two-gluon amplitudes. 
We thank S.~Badger, C.~Br{\o}nnum-Hansen, H.~B.~Hartanto and 
T.~Peraro for 
correspondence concerning the numerical results presented 
in ref.~\cite{Badger:2018gip}.
We thank C.~Duhr for the use of his \Mathematica{} package 
\texttt{PolyLogTools}.
The work
of S.A. and F.F.C. is supported by the Alexander von Humboldt 
Foundation, in the framework of the Sofja Kovalevskaja Award
2014, endowed by the German Federal
Ministry of Education and Research. 
V.S.'s work is funded by the German Research Foundation (DFG) 
within the Research Training Group GRK~2044.
F.F.C. thanks the Mainz Institute for Theoretical Physics 
(MITP) for its hospitality and partial support during the
completion of this work.
The authors acknowledge support by the state of 
Baden-W\"urttemberg through bwHPC.

\appendix
% !TEX root = ../main.tex

\section{Operations on \texorpdfstring{$\gamma$}{Gamma} Matrices}\label{sec:identities}

In the following we derive the values of the contraction of the
tensors $w_0$ and $v_n$ which are used in 
eqs.~\eqref{eq:decompqqbar} and \eqref{eqn:4qampltensor}.
In this appendix we take $d$ to be an even integer denoting the
dimension of the space for which the Clifford algebra is defined
and we denote the dimension of the $\gamma$-matrix
representation by
$d_t=\Tr(\mathbb{1}_{[d]})=2^{d/2}$.
In the main text, we are interested in the case
\bea
d = (D_s-4)\,.
\eea
Since amplitude computations are homogeneous in the
factor $d_t$, it can be factored out and replaced by 
a suitable value in order to suit the four-dimensional 
limit. In this appendix, we keep the parameter $d_t$ in analytic
form in order to maintain a consistent finite-dimensional
algebra and for clarity of the equations. 
In the main text we use formulas with the replacement
$d_t\rightarrow 1$ imposed, which is the value consistent with a
calculation in dimensional regularization \cite{Collins:1984xc}.

We start with the trivial case of $w_0$ which appears in 
eq.~\eqref{eq:decompqqbar}. It is easy to find that
\bea
w_0=\delta_\kappa^\lambda\,,\qquad 
w^0=\delta^\kappa_\lambda/d_t\,,\qquad 
w_0\cdot w^0=\delta_\kappa^\lambda  \delta^\kappa_\lambda/d_t=1\,.
\eea

For the tensors $v_n$ of eq.~\eqref{eqn:4qampltensor} we
must first consider traces of $\gamma$-matrix chains of
the form
\begin{align}\label{eq:basisGammaChainII}
\gamma_{[d]}^{\mu_1 \ldots \mu_n} = \frac{1}{n!} \sum_{ \sigma\in
S_n} \sgn(\sigma) \gamma_{[d]}^{\mu_{\sigma(1)}} \ldots
\gamma_{[d]}^{\mu_{\sigma_n}}\,,
\end{align}
with $S_n$ denoting the set of permutations of $n$
integers and $\sgn(\sigma)$ the signature of the permutation $\sigma\in S_n$.
Given a unitary
representation of the $\gamma_{[d]}^\mu$ matrices, hermitian
conjugation reverses the $\gamma$-matrix chains and flips the
Lorentz index position.
This can be seen from the definition of the Clifford algebra
(\ref{eqn:defClifford}) which implies 
$\gamma_{[d]}^\mu \gamma^{}_{[d]\mu} = \mathbb{1}_{[d]}$ for 
any fixed~$\mu$.
Assuming that the $\gamma_{[d]}^\mu$ are unitary, i.e.
$(\gamma_{[d]}^\mu)^\dagger = (\gamma_{[d]}^\mu)^{-1}$,
then implies
$(\gamma_{[d]}^\mu)^\dagger=\gamma_{[d]\mu}$. 
For the above product of $\gamma$ matrices this in turn leads to
\bea (\gamma_{[d]}^{\mu_1 \ldots \mu_n} )^\dagger =\,
\gamma_{[d]\mu_n \ldots \mu_1}^{\phantom{\mu}} \,. \eea
Unitary representations for the Clifford algebra can always be
found as explained for example in 
ref.~\cite{Kreuzer:susylectures}.
We will require the following traces of antisymmetric
$\gamma$-matrix chains,
\bea \label{eq:traceChain}
\Tr(\gamma_{[d]}^{\mu_1 \ldots \mu_n} \gamma^{
\phantom{\mu}}_{[d]\,\nu_m \ldots \nu_1})= 
\left\{ \begin{array}{cc} d_t 
\sum_{ \sigma\in  S_n} \sgn(\sigma)
\delta^{\mu_{\sigma(1)}}_{\nu_1}\cdots\delta^{\mu_{\sigma(n)}}_
{\nu_n} &\qquad m=n   \\
 0 &\qquad m\neq n\end{array}  
 \right. 
\,,
\eea
where the summation runs over all permutations $S_n$ of $n$ elements.
The traces are computed in fixed integer dimensions where the dimensions of the
$\gamma$-matrix representation is taken to be $d_t$ dimensional.
For contracted Lorentz indices we will also use that
\bea \label{eqn:simpletrace}
\sum_{\mu_1,\ldots,\mu_n} 
\sum_{ \sigma\in  S_n} \sgn(\sigma)
\delta^{\mu_{\sigma(1)}}_{\mu_1}\ldots
\delta^{\mu_{\sigma(n)}}_{\mu_n}
= \frac{d!}{(d-n)! }\,.
\eea
The sum counts the number of antisymmetric tensors of
rank $n$ in $d$ dimensions, which is the number of ways to 
choose an ordered subset of $n$ elements from a fixed set of $d$
elements.
With these preparatory equations we can
compute the inner products of the $v_n$ tensors of 
\eqn{eqn:4qtensors}
which yield the normalisation factors $c_n$
of \eqn{eqn:vnproducts}:
\begin{align}\begin{split}
\label{eq:cnCalc}
c_n=v_n^\dagger\cdot v_n =&\,
\Tr(\gamma_{[d]\,\mu_n \ldots \mu_1} \gamma_{[d]}^
{\nu_1 \ldots \nu_n}) \,
\Tr(\gamma_{[d]}^{\mu_n \ldots \mu_1} \gamma_{
[d]\,\nu_1 \ldots \nu_n}) \\
=&\,
d_t^2 \sum_{\sigma\in  S_n} 
\sum_{\mu_1,\ldots,\mu_n}
\sum_{\tilde \sigma\in S_n}
\sum_{\nu_1,\ldots,\nu_n}
 \sgn(\sigma)
 \sgn(\tilde\sigma)
\delta^{\mu_{\sigma(n)}}_{\nu_1}\cdots\delta^{\mu_{\sigma(1)}}_{\nu_n}
\delta_{\mu_{\tilde\sigma(n)}}^{\nu_1}\cdots
\delta_{\mu_{\tilde\sigma(1)}}^{\nu_n}
\\
=&\,d_t^2 \sum_{\sigma\in  S_n} 
\sgn(\sigma)
\left(
\sum_{\mu_1,\ldots,\mu_n}\sum_{\tilde \sigma\in S_n}
 \sgn(\tilde\sigma)
 \delta^{\mu_{\sigma(n)}}_{\mu_{\tilde\sigma(n)}}
 \cdots
 \delta^{\mu_{\sigma(1)}}_{\mu_{\tilde\sigma(1)}}
 \right)
\\
=&\, d_t^2 \sum_{\sigma\in  S_n} 
\sgn(\sigma)^2 \frac{d!}{(d-n)!}\\
=&\, d_t^2 \frac{d!\,n!}{(d-n)!}\,.
\end{split}\end{align}
In the above formulas the summation over the indices $\nu_i$ is 
trivially performed. In the next step, we isolate a contribution
of the form that we computed in eq.~\eqref{eqn:simpletrace},
which gives the same result for each permutation $\sigma$ but
multiplied by $\sgn(\sigma)$. The final results follows
trivially.
As expected, for each $n$ the result has zeros in the dimensions
$d$ for which there are insufficient distinct labels 
$\mu_i$ and $\nu_i$ available to form antisymmetric index 
configurations of $n$ indices.
We recall that in the main text we set $d_t=1$ and $d=D_s-4$.

Finally we collect the results for the contractions of the 
tensor $\tilde v_m$ and $v_n$ required for amplitudes with two
quark lines of identical flavor,
see eq.~\eqref{eq:basisIdentical}.
These contractions lead to a single trace instead of a product 
of traces as was the case in eq.~\eqref{eq:cnCalc}.
We refer to the above intuitive argument: tensor 
contractions including $v_n$ or $\tilde v_n$ vanish 
whenever the dimensionality $d$ is insufficient to accommodate
the respective antisymmetric index arrangements in the Lorentz 
indices $\mu_i$ and $\nu_i$.
In particular this implies that contractions including the
tensors $v_{n}$ are proportional to $d$ for $n\neq0$. We find
that 
\begin{align}\begin{split}
\tilde v_0^\dagger \cdot v_0 &=
\delta^{\kappa_1}_{\lambda_2} \delta^{\kappa_2}_{\lambda_1}
\delta_{\kappa_1}^{\lambda_1} \delta_{\kappa_2}^{\lambda_2} 
= d_t \,, \\
\tilde v_m^\dagger \cdot v_{n} &=d_t\,d\,p_{mn}(d) = 
{\cal O}(\epsilon)\,,\qquad\textrm{for }\{m,n\}\neq\{0,0\}\,.
\end{split}\end{align}
Here $p_{mn}(d)$ is a polynomial-valued matrix which we 
will not require explicitly for the present paper, and in the
last equality we made explicit the fact that in this paper we 
are interested in the case $d=D_s-4={\cal O}(\epsilon)$.

% !TEX root = ../main.tex

\section{Divergence Structure of Two-Loop Five-Parton Amplitudes}
\label{sec:IR}

We use the HV dimensional regularization
scheme to handle both ultraviolet and infrared divergences.
UV divergences are removed through renormalization and the
remaining infrared poles can be computed from the corresponding
lower-order
amplitudes~\cite{Catani:1998bh,Sterman:2002qn,Becher:2009cu,
Gardi:2009qi}. In this appendix we detail this procedure.
Reproducing the pole structure of the amplitudes we have 
computed is an important check of our results.

\subsection{Renormalization}

We perform renormalization of the QCD coupling in the 
$\overline{\text{MS}}$ scheme. It is implemented by replacing 
the bare coupling by the renormalized one, denoted $\alpha_s$, 
in eq.~\eqref{eq:partials}.
The bare and renormalized couplings are related through
\begin{equation}
    \alpha_0\mu_0^{2\epsilon}S_{\epsilon}
  =\alpha_s\mu^{2\epsilon}\left(
  1-\frac{\beta_0}{\epsilon}\frac{\alpha_s}{4\pi}
  +\left(\frac{\beta_0^2}{\epsilon^2}-\frac{\beta_1}{\epsilon}\right)
  \left(\frac{\alpha_s}{4\pi}\right)^2+\mathcal{O}\left(\alpha_s^3\right)\right)\,,
\end{equation}
where $S_\epsilon=(4\pi)^{\eps}e^{-\eps\gamma_E}$, with
$\gamma_E=-\Gamma'(1)$ the Euler-Mascheroni constant.
$\mu_0^2$ is the scale introduced in dimensional regularization
to keep the coupling dimensionless in the QCD Lagrangian, 
and $\mu^2$ is the renormalization scale. In the following, we
set $\mu_0^2=\mu^2=1$. The leading-color coefficients of the QCD
$\beta$-function are
\begin{equation}
  \beta_0=\frac{\NC}{3} \left( 11 - 2\frac{N_f}{\NC}
  \right),\qquad
  \beta_1=\frac{\NC^2}{3} \left( 17 - \frac{13}{2} \frac{N_f}{\NC} \right).
\end{equation}
The perturbative expansion of the renormalized amplitude is
\begin{equation}\label{eq:renormAmp}
  \mathcal{A}_R = S_\epsilon^{-\frac{\lambda}{2}}
  g_s^\lambda\left(
  \mathcal{A}_R^{(0)}
  +\frac{\alpha_s}{4\pi}\NC\,\mathcal{A}_R^{(1)}
  +\left(\frac{\alpha_s}{4\pi}\right)^2\NC^2\mathcal{A}_R^{(2)}
  +\mathcal{O}(\alpha_s^3)
  \right),
\end{equation}
where $\lambda$ is the power of $g_0$ in the tree amplitude, 
with $\alpha_0=g_0^2/(4\pi)$ and similarly for $\alpha_s$.
For four-parton amplitudes $\lambda=2$, and for five-parton 
amplitudes $\lambda=3$.
The renormalized amplitudes $\mathcal{A}_R^{(i)}$ are related 
to the bare amplitudes $\mathcal{A}^{(i)}$ as follows:
\begin{align}\begin{split}
  \label{eq:twoLoopUnRenorm}
    &\mathcal{A}_R^{(0)}=\mathcal{A}^{(0)}, \\
  & \mathcal{A}_R^{(1)}=S_{\epsilon}^{-1}\mathcal{A}^{(1)}
  -\frac{\lambda}{2\epsilon}\frac{\beta_0}{\NC}
  \mathcal{A}^{(0)}\,,\\
  &\mathcal{A}_R^{(2)}=
  S_{\epsilon}^{-2}\mathcal{A}^{(2)}
  -\frac{\lambda+2}{2\epsilon}\frac{\beta_0}{\NC}S_{\epsilon}^
  {-1}
  \mathcal{A}^{(1)}
  +\left(\frac{\lambda(\lambda+2)}{8\epsilon^2}\left(\frac{\beta_0}
  {\NC}\right)^2
  -\frac{\lambda}{2\epsilon}\frac{\beta_1}{\NC^2}\right)
  \mathcal{A}^{
  (0)}\,.
\end{split}\end{align}

\subsection{Infrared Behavior}

The poles of renormalized amplitudes are of infrared origin and
can be predicted from the previous orders in the perturbative 
expansion 
\cite{Catani:1998bh,Sterman:2002qn,Becher:2009cu,Gardi:2009qi}:
\begin{align}\begin{split}\label{eq:catani}
    A_R^{(1)}&={\bf I}^{(1)}_{[n]}(\epsilon)
    A_R^{(0)}+\mathcal{O}
    (\epsilon^0)\,,\\
    A_R^{(2)}&={\bf I}^{(2)}_{[n]}(\epsilon)A_R^{(0)}+{\bf I}^{(1)}_{[n]}(\epsilon)
    A_R^{(1)}+\mathcal{O}(\epsilon^0)\,,
\end{split}\end{align}
with the operators ${\bf I}^{(1)}_{[n]}$ and
${\bf I}^{(2)}_{[n]}$ depending on the number and the type of
the scattering particles. This dependence is denoted by the 
subscript $[n]$.
For amplitudes in the leading-color approximation and for which
all quark lines have distinct flavor, the operators
$\mathbf{I}^{(1)}_{[n]}$ and $\mathbf{I}^{(2)}_{[n]}$ are 
diagonal in color space and can be written in a very compact
form. The operator $\mathbf{I}^{(1)}_{[n]}$ is given by
\begin{equation}
  {\bf I}^{(1)}_{[n]}(\eps)=-\frac{e^{\gamma_E\eps}}{\Gamma(1-\epsilon)}
  \sum_{i=1}^n\gamma_{a_i,a_{i+1}}
  \left( -s_{i,i+1}\right)^{-\epsilon}\,,
\end{equation}
with the indices defined cyclically.
The index $a_i$ denotes a type of particle with momentum $p_i$, i.e., in the context of our paper,
$a_i\in\{g,q,\bar q, Q, \bar Q\}$. We introduced the auxiliary symbols $\gamma_{a,b}$, 
symmetric under the exchange of indices, 
$\gamma_{a,b}=\gamma_{b,a}$, and defined according to:
\begin{align}\begin{split}
  \gamma_{g,g}&=\frac{1}{\epsilon^2}+\frac{1}{2\eps}
  \frac{\beta_0}{\NC}\,, \\
  \gamma_{q,Q}&=\gamma_{q,\bar Q}=
  \gamma_{\bar q, Q}=\gamma_{\bar q, \bar Q} 
  =\frac{1}{\epsilon^2}+\frac{3}{2\eps}\,,\\
  \gamma_{g,q}&=\gamma_{g,\bar q}=
  \gamma_{g,Q}=\gamma_{g,\bar Q}=
  \frac{\gamma_{g,g}+\gamma_{q,Q}}{2}\,,\\
  \gamma_{q,\bar q}&=\gamma_{Q,\bar Q}=0\,.
\end{split}\end{align}
The operator~${\bf I}^{(2)}_{[n]}$ is
\begin{align}\begin{split} \label{eqn:Iop}
		{\bf I}^{(2)}_{[n]}(\eps)=&
  -\frac{1}{2}{\bf I}^{(1)}_{[n]}(\eps){\bf I}^{(1)}_{[n]}(\eps)
  -\frac{\beta_0}{\NC\epsilon}{\bf I}^{(1)}_{[n]}(\eps) + 
	\frac{e^{-\gamma_E\epsilon}\Gamma(1-2\epsilon)}
  {\Gamma(1-\epsilon)}
  \left(\frac{\beta_0}{\NC\epsilon}+K\right)
  {\bf I}^{(1)}_{[n]}(2\epsilon) + 
  {\bf H}_{[n]}(\epsilon)\,,
\end{split}\end{align}
where 
\begin{equation}
K=\frac{67}{9}-\frac{\pi ^2}{3}-\frac{10}{9}\frac{\NF}{\NC}\,,
\end{equation}
and ${\bf H}_{[n]}(\epsilon)$ is a diagonal operator at 
leading color that depends on the number of external quarks 
and gluons in the process,
\begin{align}\begin{split}
  {\bf H}_{[n]}(\epsilon)&=
  \frac{e^{\gamma_E\epsilon}}{\epsilon\Gamma(1-\epsilon)}
  \sum_{i=1}^n\left(
  \delta_{a_i,g}H_g+
  (\delta_{a_i,q}+\delta_{a_i,\bar q}
  +\delta_{a_i,Q}+\delta_{a_i,\bar Q})
  H_q
  \right)\,,
\end{split}\end{align}
with (see e.g.~\cite{Bern:2003ck})
\begin{align}\begin{split}
  H_g&= \left(\frac{\zeta_3}{2}+\frac{5}{12}+
  \frac{11\pi^2}{144}\right)
  -\left(\frac{\pi^2}{72}+\frac{89}{108}\right)\frac{N_f}{\NC}
  +\frac{5}{27}\left(\frac{N_f}{\NC}\right)^2\,,\\
  H_q&=
  \left(\frac{7\zeta_3}{4}+\frac{409}{864}
  -\frac{11\pi^2}{96}\right)
  +\left(\frac{\pi^2}{48}-\frac{25}{216}\right)\frac{N_f}{\NC}\,.
\end{split}\end{align}

The poles of the bare amplitudes, as presented for example in
tables~\ref{tab:results4parton} and \ref{tab:results5parton}, can be recovered
from those of the renormalized amplitude by using
eqs.~\eqref{eq:twoLoopUnRenorm}.

\subsection{Numerical Results for One-loop Amplitudes}
\label{sec:oneloopvalues}

To predict the expected pole structure of the amplitudes
computed in section~\ref{sec:Results} it is necessary to compute
corresponding one-loop results up to high enough order in
$\epsilon$. For completeness, we present one-loop
results in tables \ref{tab:results4parton1L} and 
\ref{tab:results5parton1L} which we have obtained with our 
own implementation of one-loop numerical unitarity. 
The expansion has been performed up to 
${\cal O}(\epsilon^2)$ in order to allow the evaluation of 
finite remainders as in eq.~\eqref{eq:remainderDef}.
This was used to reproduce the analytic results for
finite remainders of the $q\bar q gg$ and $q\bar q Q\bar Q$
amplitudes of refs.~\cite{Bern:2003ck,Glover:2004si}. 
The results are normalized to remove overall phase ambiguities 
in the amplitudes, choosing the tree-level amplitude if it does
not vanish, or the leading term of the one-loop amplitude
otherwise. We present numerical values with 10 significant
digits.

\begin{table}[h]
  \centering
  \begin{adjustbox}{width=1\textwidth}
    \begin{tabular}{cccccc}
      \toprule
      $\CA^{(1)[N_f^0]}/\CA^{\rm (norm)}$   &   $\epsilon^{-2}$   &   $\epsilon^{-1}$   &   $\epsilon^{0}$   &   $\epsilon^{1}$  &  $\epsilon^{2}$ \\
      \midrule
      $(1_g^+,2_g^+,3_g^+,4_g^+)$ & $0$ & $0$ & $1$ & 
      $3.144383516$
      & $4.993655130$ \\
      $(1_g^-,2_g^+,3_g^+,4_g^+)$ & $0$ & $0$ & $1$ & 
      $6.037021519$
      & $19.41121185$ \\
      $(1_g^-,2_g^-,3_g^+,4_g^+)$ &  $-4.000000000$ &
      $-14.82985386$ &
      $-21.50563510$ & $-4.242972632$ & $39.45669987$ \\
      $(1_g^-,2_g^+,3_g^-,4_g^+)$ &  $-4.000000000$ &
      $-14.82985386$ & $-24.34737636$ & $-23.80446527$ &
      $-30.91926414$\\ \midrule
      $(1_q^+,2_{\bar q}^-,3_g^+,4_g^+)$ & $0$ & $0$ & $1$ &
      $5.886473216$  & $18.18093693$   \\
      $(1_q^+,2_{\bar q}^-,3_g^+,4_g^-)$ & $-3.000000000$ &
      $-10.42169654$ & $-13.75537910$ & $-2.227311547$ &
      $15.67564907$ \\
      $(1_q^+,2_{\bar q}^-,3_g^-,4_g^+)$ & $-3.000000000$ &
      $-10.42169654$ &
      $-10.64041688$ & $20.52306512$ & $101.8467214$ \\
      \midrule
      $(1_q^+,2_{\bar q}^-,3_Q^-,4_{\bar Q}^+)$ & $-2.000000000$ &
      $-6.013539220$  & $4.503971305$ & $55.27734017$ &
      $156.3375209$ \\
      $(1_q^+,2_{\bar q}^-,3_Q^+,4_{\bar Q}^-)$ & $-2.000000000$ &
      $-6.013539220$  & $-1.512122300$ & $22.96961380$ &
      $57.55706218$ \\
      \toprule
      $\CA^{(1)[N_f^1]}/\CA^{\rm (norm)}$   &   $\epsilon^{-2}$   &   $\epsilon^{-1}$   &   $\epsilon^{0}$   &   $\epsilon^{1}$  &  $\epsilon^{2}$ \\
      \midrule
      $(1_g^+,2_g^+,3_g^+,4_g^+)$ & $0$ & $0$ & $-1.000000000$ &
      $-4.144383516$ & $-9.138038646$\\
      $(1_g^-,2_g^+,3_g^+,4_g^+)$ & $0$ & $0$ & $-1.000000000$ &
      $-7.037021519$ & $-26.44823337$ \\
      $(1_g^-,2_g^-,3_g^+,4_g^+)$ & $0$ & $0.6666666667$ &
      $3.337846407$ & $7.778113386$ & $9.642499788$ \\
      $(1_g^-,2_g^+,3_g^-,4_g^+)$ & $0$ & $0.6666666667$ &
      $3.888266255$ & $11.57993010$ & $23.40355137$ \\
      \midrule
      $(1_q^+,2_{\bar q}^-,3_g^+,4_g^+)$ & $0$ & $0$ &
      $-0.1818181818$ & $-1.074210422$ & $-3.518712119$ \\
      $(1_q^+,2_{\bar q}^-,3_g^+,4_g^-)$ & $0$ & $0$ & $0$ & $0$ & $0$ \\
      $(1_q^+,2_{\bar q}^-,3_g^-,4_g^+)$ & $0$ & $0$ & $0$ & $0$ & $0$ \\
      \midrule
      $(1_q^+,2_{\bar q}^-,3_Q^-,4_{\bar Q}^+)$ & $0$ &
      $-0.6666666667$ & $-2.605438214$ & $-5.691068008$ &
      $-8.728233619$ \\
      $(1_q^+,2_{\bar q}^-,3_Q^+,4_{\bar Q}^-)$ & $0$ &
      $-0.6666666667$ & $-2.605438214$ & $-5.691068008$ &
      $-8.728233619$ \\
      \bottomrule
    \end{tabular}
  \end{adjustbox}
\caption{The bare one-loop four-parton helicity amplitudes 
  evaluated at the phase space point in 
  eq.~\eqref{eq:EvalPoint4}. We set the
    normalization factor $\CA^{\rm (norm)}$ to $\CA^{(1)[N_f^0]}(\epsilon=0)$ for the
    amplitudes with vanishing trees, and to $\CA^{(0)}$ otherwise.}
  \label{tab:results4parton1L}
\end{table}

\begin{table}[h]
  \centering
  \begin{adjustbox}{width=1\textwidth}
    \begin{tabular}{cccccc}
      \toprule
      $\CA^{(1)[N_f^0]}/\CA^{\rm (norm)}$   &   $\epsilon^{-2}$   &   $\epsilon^{-1}$   &   $\epsilon^{0}$   &   $\epsilon^{1}$  &  $\epsilon^{2}$ \\
      \midrule
      $(1_g^+,2_g^+,3_g^+,4_g^+,5_g^+)$ & $0$ & $0$ & $1$ &
      $3.033832975$ & $4.587604357$ \\
      $(1_g^-,2_g^+,3_g^+,4_g^+,5_g^+)$ & $0$ & $0$ & $1$ &
      $5.624431423$ & $16.89796219$ \\
      $(1_g^-,2_g^-,3_g^+,4_g^+,5_g^+)$ &  $-5.000000000$ &
      $-17.88291386$ & $-24.30905600$ & $0.2206218531$ &
      $59.35260478$ \\
      $(1_g^-,2_g^+,3_g^-,4_g^+,5_g^+)$ &  $-5.000000000$ &
      $-17.88291386$ & $-29.50855173$ & $-34.92963561$ &
      $-64.50302993$ \\
      \midrule
      $(1_q^+,2_{\bar q}^-,3_g^+,4_g^+,5_g^+)$ & $0$ & $0$ & $1$ &
      $5.892137144$ & $18.35590938$ \\
      $(1_q^+,2_{\bar q}^-,3_g^+,4_g^+,5_g^-)$ & $-4.000000000$ &
      $-13.76243861$ & $-15.50477253$ & $17.23285932$ &
      $101.5375461$ \\
      $(1_q^+,2_{\bar q}^-,3_g^+,4_g^-,5_g^+)$ & $-4.000000000$ &
      $-13.76243861$ & $-17.97203103$ & $1.496892271$ &
      $50.75427433$ \\
      $(1_q^+,2_{\bar q}^-,3_g^-,4_g^+,5_g^+)$ & $-4.000000000$ &
      $-13.76243861$ & $-16.98218729$ & $7.025105072$ &
      $65.53899984$ \\
      \midrule
      $(1_q^+,2_{\bar q}^-,3_Q^-,4_{\bar Q}^+,5_g^+)$ &
      $-3.000000000$& $-8.843501370$ & $-1.852152501$ &
      $37.28945738$
      & $105.9935237$ \\
      $(1_q^+,2_{\bar q}^-,3_Q^+,4_{\bar Q}^-,5_g^+)$ &
      $-3.000000000$& $-8.843501370$ & $-4.411871382$ &
      $26.32328221$ & $81.15715418$ \\
      $(1_q^+,2_{\bar q}^-,3_Q^-,4_{\bar Q}^+,5_g^-)$ &
      $-3.000000000$& $-8.843501370$ & $342.9945174$ &
      $1000.539160$ & $-355.3299610$ \\
      $(1_q^+,2_{\bar q}^-,3_Q^+,4_{\bar Q}^-,5_g^-)$ &
      $-3.000000000$&
      $-8.843501370$ & $-1.744812968$ & $-9.470771643$ &
      $-176.4533405$ \\
      \toprule
      $\CA^{(1)[N_f^1]}/\CA^{\rm (norm)}$   &   $\epsilon^{-2}$   &   $\epsilon^{-1}$   &   $\epsilon^{0}$   &   $\epsilon^{1}$  &  $\epsilon^{2}$ \\
      \midrule
      $(1_g^+,2_g^+,3_g^+,4_g^+,5_g^+)$ & $0$ & $0$ &
      $-1.000000000$ & $-4.033832975$ & $-8.621437332$ \\
      $(1_g^-,2_g^+,3_g^+,4_g^+,5_g^+)$ & $0$ & $0$ &
      $-1.000000000$ & $-6.624431423$ & $-23.52239361$ \\
      $(1_g^-,2_g^-,3_g^+,4_g^+,5_g^+)$ &  $0$ & $0.6666666667$ &
      $2.494683591$ & $2.329188091$ & $-8.735477566$ \\
      $(1_g^-,2_g^+,3_g^-,4_g^+,5_g^+)$ &  $0$ & $0.6666666667$ &
      $3.475701080$ &  $8.982161551$ &  $14.85398827$ \\
      \midrule
      $(1_q^+,2_{\bar q}^-,3_g^+,4_g^+,5_g^+)$ & $0$ & $0$ &
      $-0.3542206031$ & $-2.268220888$ & $-7.918667025$ \\
      $(1_q^+,2_{\bar q}^-,3_g^+,4_g^+,5_g^-)$ & $0$ & $0$ &
      $-0.5535785746$ & $-3.637432164$ & $-12.69744845$ \\
      $(1_q^+,2_{\bar q}^-,3_g^+,4_g^-,5_g^+)$ & $0$ & $0$ &
      $-0.0008015520164$ & $-0.004344237791$ & $-0.01257682159$
      \\
      $(1_q^+,2_{\bar q}^-,3_g^-,4_g^+,5_g^+)$ & $0$ & $0$ &
      $-0.04501904941$ & $-0.2962279378$ & $-1.036895298$ \\
      \midrule
      $(1_q^+,2_{\bar q}^-,3_Q^-,4_{\bar Q}^+,5_g^+)$ & $0$ &
      $-0.6666666667$ & $-2.939327989$ & $-7.089932089$ &
      $-11.96893214$ 
      \\
      $(1_q^+,2_{\bar q}^-,3_Q^+,4_{\bar Q}^-,5_g^+)$ & $0$ &
      $-0.6666666667$ & $-2.933154494$ & $-7.055606900$ &
      $-11.86563786$ 
      \\
      $(1_q^+,2_{\bar q}^-,3_Q^-,4_{\bar Q}^+,5_g^-)$ & $0$ &
      $-0.6666666667$ & $-57.49865762$ & $-259.2491530$ &
      $-668.4609808$
      \\
      $(1_q^+,2_{\bar q}^-,3_Q^+,4_{\bar Q}^-,5_g^-)$ & $0$ &
      $-0.6666666667$ & $1.491351996$ & $9.944256190$ & 
      $24.03526126$ \\
      \bottomrule
    \end{tabular}
  \end{adjustbox}
  \caption{The bare one-loop five-parton helicity amplitudes 
  evaluated at the phase space point in 
  eq.~\eqref{eq:EvalPoint5}. We set the
    normalization factor $\CA^{\rm (norm)}$ to $\CA^{(1)[N_f^0]}(\epsilon=0)$ for the
    amplitudes with vanishing trees, and to $\CA^{(0)}$ otherwise.}
  \label{tab:results5parton1L}
\end{table}

\newpage

\bibliography{main.bib}

\end{document}